\documentstyle[epsfig,12pt,a4p]{article}
 
\newcommand{\ra}{\mbox{$\rightarrow \;$}}

\newcommand{\etal}{\mbox{\it et al.}}
\newcommand {\beq}{\begin{equation}}
\newcommand {\eeq}{\end{equation}}
\newcommand {\bea}{\begin{eqnarray}}
\newcommand {\eea}{\end{eqnarray}}

\newcommand {\SM}{Standard Model}

\newcommand{\Z}{\mbox{\rm Z$^0$ }}
\newcommand{\tautau}{\mbox{\tp \tm}}
\newcommand{\ee}{\mbox{\elp \elm} }
\newcommand{\mumu}{\mbox{\mup \mum} }

\newcommand{\qq}{\mbox{$\rm q \bar{q}$}}

\newcommand{\tn}{\mbox{$\tau$}}
\newcommand{\tm}{\mbox{$\tau^-$}}
\newcommand{\tp}{\mbox{$\tau^+$}}
\newcommand{\nut}{\mbox{$\nu_{\tau}$}}

\newcommand{\eln}{\mbox{\rm e}}
\newcommand{\elm}{\mbox{\rm e$^-$}}
\newcommand{\elp}{\mbox{\rm e$^+$}}
\newcommand{\nue}{\mbox{$\nu_e$}}

\newcommand{\nueb}{\mbox{$\bar{\nu}_e$}}
 
\newcommand{\mun}{\mbox{$\mu$}}
\newcommand{\mum}{\mbox{$\mu^-$}}
\newcommand{\mup}{\mbox{$\mu^+$}}
\newcommand{\num}{\mbox{$\nu_{\mu}$}}

\newcommand{\pin}{\mbox{$\pi$}}
\newcommand{\pim}{\mbox{$\pi^-$}}
\newcommand{\pip}{\mbox{$\pi^+$}}
\newcommand{\piz}{\mbox{$\pi^0$}}
 
\newcommand{\piK}{\mbox{$\pi$(K)}}
 
\newcommand{\rhon}{\mbox{$\rho$}}

\newcommand{\ksh}{\mbox{\rm K$^0_{\mathrm{S}}$}}

\newcommand{\aone}{\mbox{\rm a$_1$}}

\newcommand{\eeee}{\mbox{$\ee\ra\ee$} }
\newcommand{\eemm}{\mbox{$\ee\ra\mumu$} }
\newcommand{\eett}{\mbox{$\ee\ra\tautau$} }
\newcommand{\eeeeee}{\mbox{$\ee\ra\ee\ee$} }
\newcommand{\eeeemm}{\mbox{$\ee\ra\ee\mumu$} }

\newcommand{\tmu}{\mbox{\tn \ra \mun \num \nut}}
\newcommand{\tel}{\mbox{\tn \ra \eln \nue \nut}}

\newcommand{\tpi}{\mbox{\tn \ra \pin\nut}}
\newcommand{\tK}{\mbox{\tn \ra K \nut}}

\newcommand{\tro}{\mbox{\tn \ra \rhon\nut}}
\newcommand{\ttwopi}{\mbox{\tn \ra $\pi \pi^0$ \nut}}
\newcommand{\tKpiz}{\mbox{\tn \ra K $\pi^0$ \nut}}

\newcommand{\taone}{\mbox{\tn \ra a$_1$\nut}}
\newcommand{\threepi}{\mbox{\tn \ra 3$\pi^{\pm}$ \nut}}
\newcommand{\Kpipi}{\mbox{\tn \ra K $\pi^+\pi^-$ \nut}}

\newcommand{\thnpiz}{\mbox{\tn \ra \piK $\geq$2\piz  \nut}}
\newcommand{\tpitwopiz}{\mbox{\tn \ra $\pi 2 \piz$  \nut}}

\newcommand{\efswsq}
{\mbox{$\sin^2\theta_{\mbox{\scriptsize\rm eff}}
                    ^{\mbox{\scriptsize\rm lept}}$}}
\newcommand{\Atau}{\mbox{$\cal A_{\tau}$}}
\newcommand{\Amu}{\mbox{$\cal A_{\mu}$}}
\newcommand{\Aele}{\mbox{$\cal A_{\mathrm e}$}}
\newcommand{\Alep}{\mbox{$\cal A_{\ell}$}}

\newcommand{\gvt}{\mbox{$g_{\mathrm{V\tau}}$}}
\newcommand{\gat}{\mbox{$g_{\mathrm{A\tau}}$}}
\newcommand{\gve}{\mbox{$g_{\mathrm{Ve}}$}}
\newcommand{\gae}{\mbox{$g_{\mathrm{Ae}}$}}
\newcommand{\gvm}{\mbox{$g_{\mathrm{V\mu}}$}}
\newcommand{\gam}{\mbox{$g_{\mathrm{A\mu}}$}}
\newcommand{\gvl}{\mbox{$g_{\mathrm{V\ell}}$}}
\newcommand{\gal}{\mbox{$g_{\mathrm{A\ell}}$}}

\newcommand{\vovat}{\mbox{\gvt/\gat}}
\newcommand{\vovae}{\mbox{\gve/\gae}}
\newcommand{\voval}{\mbox{\gvl/\gal}}

\newcommand{\act}{\mbox{$|\cos\theta_{\mathrm{jet}}|$}}
\newcommand{\cst}{\mbox{$\cos\theta_{\tau^-}$}}

\newcommand{\cstsq}{\mbox{$\cos^{2}\theta_{\tau^-}$}}

\newcommand{\ecm}{\mbox{$\sqrt{\mathrm{s}}$}}
\newcommand{\mz}{{M$_{\mathrm{Z}}$}}

\newcommand{\MZ}{{M_{\mathrm{Z}}}}
\newcommand {\MH}         {m_{\mathrm{H }}}
\newcommand {\Mtop}         {m_{\mathrm{t}}}
\newcommand {\als}          {\alpha_{\mathrm{s}}}

\newcommand {\GZ}         {\Gamma_{\mathrm{Z}}}
\newcommand {\shadpol}    {\sigma^0_{\mathrm{h}}}
\newcommand {\Ree}        {R_{\mathrm{e}}}
\newcommand {\Rmu}        {R_{\mu}}
\newcommand {\Rtau}       {R_{\tau}}
\newcommand {\Afbpolee}   {A_{\mathrm{FB}}^{0,\mathrm{e}}}
\newcommand {\Afbpolmumu} {A_{\mathrm{FB}}^{0,\mu}}
\newcommand{\Afbpoltautau}{A_{\mathrm{FB}}^{0,\tau}}

\newcommand {\gev}        {\left( \mathrm{GeV} \right) }
\newcommand {\nb}         {\left( \mathrm{nb} \right)  }

\newcommand {\SLP}     {model-independent \Z ~parameters}

\newcommand{\ptm}{\mbox{$P_{\tau^-}$}}
\newcommand{\ptp}{\mbox{$P_{\tau^+}$}}
\newcommand{\pta}{\mbox{$\langle P_{\tau}\rangle$~}}
\newcommand{\ptau}{\mbox{$P_{\tau}$}}

\newcommand{\Pt}{\mbox{\rm P$_{\tau}$~}}
\newcommand{\afb}{\mbox{\rm A$_{\mathrm {FB}}$~}}
\newcommand{\aplfb}{\mbox{\rm A$_{\mathrm {pol}}^{\mathrm {FB}}$~}}

\newcommand{\sigp}{\mbox{$\sigma_+$}}
\newcommand{\sigm}{\mbox{$\sigma_-$}}
\newcommand{\sigtot}{\mbox{$\sigma_{total}$}}

\newcommand{\sij}{\mbox{$\sigma_{ij}$}}
\newcommand{\xxi}{\mbox{$x_i$}}
\newcommand{\xxj}{\mbox{$x_j$}}

\newcommand{\thovsxt}{\mbox{$\frac{3}{16}$}}

\newcommand{\eitovth}{\mbox{$\frac{8}{3}$}}

\newcommand{\tjet}{\mbox{\tn -jet}}
\newcommand{\tjets}{\mbox{\tn -jets}}

\newcommand{\Ntpair}{$144,810$}
\newcommand{\NtpairSel}{$129,902$}
\newcommand\Nrho{67,682}
\newcommand\brrho{0.25}
\newcommand\rhoeff{73\%}
\newcommand\rhobkg{29\%}

\newcommand\rhosensi{0.49}
 \newcommand\ptaurho{$-13.3\!\pm\! 1.1$}
 \newcommand\apolfbrho{$-10.6\!\pm\! 1.1$}

\newcommand\Npi{30,440}
\newcommand\brpi{0.12}
\newcommand\pieff{75\%}
\newcommand\pibkg{26\%}

\newcommand\pisensi{0.58}
\newcommand\ptaupi{$-13.8\! \pm \! 1.2$}
\newcommand\apolfbpi{$-11.5\! \pm \! 1.3$}

\newcommand\Nel{44,083}
\newcommand\brel{0.18}
\newcommand\eleff{92\%} 
\newcommand\elbkg{4.6\%}

\newcommand\elsensi{0.22}
\newcommand\ptauel{$-18.7\! \pm \! 2.5$}
\newcommand\apolfbel{$-8.9\! \pm \! 2.6$}

\newcommand\Nmu{41,291}
\newcommand\brmu{0.17}
\newcommand\mueff{87\%}
\newcommand\mubkg{3.3\%}

\newcommand\musensi{0.22}
\newcommand\ptaumu{$-16.3\! \pm \! 2.7$}
\newcommand\apolfbmu{$-10.6\! \pm \! 2.8$}

\newcommand\Naone{22,161}
\newcommand\braone{0.09}
\newcommand\aoneeff{77\%}
\newcommand\aonebkg{25\%}

\newcommand\aonesensi{0.45}
\newcommand\ptauaone{$-11.6\! \pm \! 2.8$}
\newcommand\apolfbaone{$-7.1\! \pm \! 2.8$}

\newcommand{\PTALL}{-14.10} 
\newcommand{\PTALLST}{0.73}  
\newcommand{\PTALLSY}{0.55}  
\newcommand{\VOVAT}{0.0732}
\newcommand{\VOVATSI}{0.0048}
\newcommand{\VOVAE}{0.0731}
\newcommand{\VOVAESI}{0.0057}
\newcommand{\APFBALL}{-10.55}  
\newcommand{\APFBALLST}{0.76}  
\newcommand{\APFBALLSY}{0.25}  
\newcommand{\AATAU}   {0.1456} 
  
\newcommand{\AATAUST} {0.0076}  
\newcommand{\AATAUSY} {0.0057}  
\newcommand{\AAEL}    {0.1454} 

\newcommand{\AAELST}  {0.0108}  
\newcommand{\AAELSY}  {0.0036}  

\newcommand{\AALEP}    {0.1455} 

\newcommand{\AALEPSIG} {0.0073}

\newcommand{\pminus}  {$89.5\pm0.2$}  
\newcommand{\peak}    {$91.25\pm0.05$}  
\newcommand{\pplus}   {$93.0\pm0.2$}  

\newcommand{\ptam}   {$-15.9\pm3.3$}  
\newcommand{\ptapk}  {$-13.52\pm0.78$}  
\newcommand{\ptap}   {$-18.1\pm2.9$}  

\newcommand{\aplfbm}   {$ -8.0\pm3.4$}  
\newcommand{\aplfbpk}  {$-10.52\pm0.81$}  
\newcommand{\aplfbp}   {$-11.3\pm3.0$}  

\newcommand{\atm}   {$0.186\pm0.039$}  
\newcommand{\atpk}  {$0.1393\pm0.0081$}  
\newcommand{\atp}   {$0.176\pm0.028$}  

\newcommand{\aem}   {$0.125\pm0.040$}  
\newcommand{\aepk}  {$0.145\pm0.011$}  
\newcommand{\aep}   {$0.146\pm0.040$}  

\newcommand{\SINWALL}{0.23172}
\newcommand{\SINWALLSI}{0.00092}
\newcommand{\SINWALLOPAL}  {0.23211}
\newcommand{\SINWALLSIOPAL}{0.00068}

\begin{document}
%
\begin{titlepage}
\begin{center}
{\large\bf EUROPEAN ORGANIZATION FOR NUCLEAR RESEARCH}
\end{center}
\begin{flushright}
CERN-EP-2001-023\\
8 March 2001\\

\end{flushright}
\vspace{3cm}
 
\begin{center}
 {\Large \bf
  Precision Neutral Current Asymmetry Parameter \\
  Measurements from the Tau Polarization at LEP}
\end{center}
\bigskip
\begin{center}
{\Large \bf The OPAL Collaboration}
\end{center}


\begin{abstract}
\noindent
Measurements of
the $\tau$ lepton polarization and forward-backward polarization
 asymmetry near the \Z ~resonance using the OPAL detector are described.
 The measurements are 
based on analyses of  \tel , \tmu , \tpi , \tro ~and \taone
~decays from a sample of ~\Ntpair ~\eett ~candidates
corresponding to an integrated luminosity of 151~pb$^{-1}$.
 Assuming that the $\tau$ lepton decays
according to V$-$A theory, we measure the average \tn ~polarization
near $\sqrt{s}$~=~\mz ~to be
$\pta= (\PTALL \pm \PTALLST \pm \PTALLSY)\%$
 and the \tn ~polarization forward-backward
asymmetry to be $\aplfb=(\APFBALL \pm \APFBALLST \pm \APFBALLSY)\%$,
where the first error is statistical and the second systematic.
Taking into account the small effects of  the photon propagator, 
photon-Z$^0$ interference and photonic radiative corrections, these results
can be expressed in terms of the lepton neutral current asymmetry parameters:
\bea
\Atau &  = & \AATAU \pm \AATAUST 
\pm \AATAUSY , \nonumber \\
\Aele & = & \AAEL \pm \AAELST 
\pm \AAELSY .  \nonumber
\eea
These measurements are
consistent with the hypothesis of lepton universality and
combine to give $\Alep = \AALEP \pm \AALEPSIG$.
 Within the context of the \SM ~this
combined result corresponds to $\efswsq=\SINWALL \pm \SINWALLSI$.
 Combing these results with those from the other OPAL neutral
 current measurements yields a value of
$\efswsq=\SINWALLOPAL \pm \SINWALLSIOPAL$.

\end{abstract}

\vfill
\begin{center}
{\large (submitted to the European Physical Journal C) } \\
\end{center}
\bigskip
\end{titlepage}

\begin{center}{\Large        The OPAL Collaboration}\end{center}\bigskip
\begin{center}{
G.\thinspace Abbiendi$^{  2}$,
C.\thinspace Ainsley$^{  5}$,
P.F.\thinspace {\AA}kesson$^{  3}$,
G.\thinspace Alexander$^{ 22}$,
J.\thinspace Allison$^{ 16}$,
G.\thinspace Anagnostou$^{  1}$,
K.J.\thinspace Anderson$^{  9}$,
S.\thinspace Arcelli$^{ 17}$,
S.\thinspace Asai$^{ 23}$,
D.\thinspace Axen$^{ 27}$
% ...
}\end{center}\bigskip
\bigskip
$^{  1}$School of Physics and Astronomy, University of Birmingham,
Birmingham B15 2TT, UK
\newline
$^{  2}$Dipartimento di Fisica dell' Universit\`a di Bologna and INFN,
I-40126 Bologna, Italy
\newline
$^{  3}$Physikalisches Institut, Universit\"at Bonn,
D-53115 Bonn, Germany
\newline
% ...
$^{  5}$Cavendish Laboratory, Cambridge CB3 0HE, UK
\newline
% ...
$^{  9}$Enrico Fermi Institute and Department of Physics,
University of Chicago, Chicago IL 60637, USA
\newline
% ...
$^{ 16}$Department of Physics, Schuster Laboratory, The University,
Manchester M13 9PL, UK
\newline
$^{ 17}$Department of Physics, University of Maryland,
College Park, MD 20742, USA
\newline
% ...
$^{ 22}$Department of Physics and Astronomy, Tel Aviv University,
Tel Aviv 69978, Israel
\newline
$^{ 23}$International Centre for Elementary Particle Physics and
Department of Physics, University of Tokyo, Tokyo 113-0033, and
Kobe University, Kobe 657-8501, Japan
\newline
% ...
$^{ 27}$University of British Columbia, Department of Physics,
Vancouver BC V6T 1Z1, Canada
\newline
% ...
Tel Aviv 69978, Israel.

\newpage

\noindent
\section{Introduction}
\label{sec-Introduction}
Parity violation in the weak neutral current 
results in a polarization of final-state fermion$-$anti-fermion pairs
produced in \Z decay with
the $\tau$ lepton being the only fundamental fermion whose polarization
 is experimentally accessible using the detectors 
at the LEP \ee collider. Consequently, measurements of the $\tau$ 
polarization provide a means for determining 
neutral current asymmetry parameters which depend on the
 neutral current vector and axial vector coupling constants.
In the \SM ~the effective electroweak mixing angle is determined from 
these couplings, therefore the polarization measurements yield a value
 of \efswsq.

 The $\tau$ polarization, \ptau,  is defined as
 $\ptau \equiv (\sigma_+ - \sigma_-)/(\sigma_+ + \sigma_-)$,
 where $\sigma_{+(-)}$ represents the cross-section for producing
 positive(negative) helicity
\tm ~leptons\footnote{By convention, $\ptau = \ptm$ and
 since, to a very good approximation, the \tm ~and \tp ~in a given event have
opposite helicities at LEP: $\ptm=-\ptp$. }. 
Assuming unpolarized \ee ~beams,
the spin-1 nature of the intermediate state implies that the
lowest order differential cross-sections for $\sigp$ and $\sigm$ 
can be expressed as:
\bea
\label{dspmdcst1}
\frac{1}{\sigtot}\frac{d\sigp}{d\cst} & = & \thovsxt
[(1+\pta)(1+\cstsq)+\eitovth(\afb+\aplfb)\cst]  \nonumber \\
\frac{1}{\sigtot}\frac{d\sigm}{d\cst} & = & \thovsxt
[(1-\pta)(1+\cstsq)+\eitovth(\afb-\aplfb)\cst]  
\eea
where 
 $\theta_{\tau^-}$ is the angle between the
 \elm ~beam and the final-state \tm ;
$\sigtot=\left[ \sigma_+ +  \sigma_- \right]_{-1<\cos\theta_{\tau^-}<1}$;
\pta ~is the average $\tau$ polarization,
 \[ \pta \equiv \frac{
               \left[ \sigma_+ \right]_{-1<\cos\theta_{\tau^-}<1} -
               \left[ \sigma_- \right]_{-1<\cos\theta_{\tau^-}<1} 
                       }
                       { \sigtot } ; \]
\afb ~is the forward-backward asymmetry of the $\tau$-pairs,
 \[ \afb \equiv \frac{
               \left[ \sigma \right]_{\cos\theta_{\tau^-}>0} -
               \left[ \sigma \right]_{\cos\theta_{\tau^-}<0} 
                       }
                       { \sigtot } ; \]
 and \aplfb ~is the forward-backward $\tau$ polarization asymmetry,
 \[ \aplfb \equiv \frac{
               \left[ \sigma_+ -  \sigma_- \right]_{\cos\theta_{\tau^-}>0} -
               \left[ \sigma_+ -  \sigma_- \right]_{\cos\theta_{\tau^-}<0} 
                       }
                       { \sigtot } \cdot \]
Equation \ref{dspmdcst1} implies a simple 
 dependence of \ptau~ on \cst :
\begin{equation}
\label{eq-ptcos}
\ptau(\cos\theta_{\tau^-})=\frac
{\pta(1+\cos^2\theta_{\tau^-}) + \frac{8}{3}\aplfb\cos\theta_{\tau^-}}
{(1+\cos^2\theta_{\tau^-}) + \frac{8}{3}\afb\cos\theta_{\tau^-}} 
\end{equation}
where the three parameters, \pta, \aplfb ~and \afb ~are extracted from data.
 These parameters include contributions from  \Z ~exchange, photon exchange 
and photon-\Z 
interference, as well as from photonic radiative corrections.
At LEP~1, where \ecm ~is near \mz, the pure \Z exchange term dominates
the polarization.  Nonetheless, the small contributions arising from the other
components must still be taken into account in the interpretation of
these parameters in the context of the neutral current couplings. The
effects of the \ecm ~dependence of the relative sizes of these contributions,
which are non-negligible near the \Z ~pole, must also be considered for a
precise neutral current interpretation.

When only the 
pure \Z ~exchange is considered, the interpretation of the measured values of
\pta ~and \aplfb in terms of the neutral current asymmetry parameters
\Atau ~and \Aele ~is remarkably simple:
\beq
\label{asymsm}
\pta=-\Atau \;\;\;\;\;\; {\mathrm {and}}
 \;\;\;\;\;\; \aplfb=-\frac{3}{4}\Aele \;\;\;\;\;\;\;\;\; \nonumber
\eeq
where the asymmetry parameters are defined\cite{bib-PCP} as:
\beq
\label{lamld}
\Alep\equiv\frac{2\voval}{1+(\voval)^2}  \nonumber
\eeq
and the symbols \gvl ~and \gal ~represent the 
effective neutral current vector and axial vector couplings for lepton 
$\ell$ as defined in reference \cite{bib-PCP}.

The inequality of the \Z coupling to left-handed and right-handed
initial-state electrons results in a polarization of the \Z 
~which manifests itself as \aplfb, whereas \pta ~expresses 
the inequality of the \Z coupling to the two chiral states of the
 final state $\tau$ leptons. Therefore,  once the small effects of 
 photon exchange,
photon-\Z ~interference and photonic radiative corrections are taken into
account,  the measurement
of \pta ~is directly related to the  ratio of the vector to axial vector
coupling constants for $\tau$ leptons and that of \aplfb ~to the ratio
  for electrons~\cite{bib-Jadach}.   Consequently, these measurements
 test the hypothesis of lepton universality in the neutral current.

In the context of the \SM,  \gvl ~and \gal ~are related to the effective
electroweak mixing angle by:
\beq
\label{vovad}
\voval=1-4~\efswsq. \nonumber
\eeq
Therefore \pta ~and \aplfb ~also provide a precision determination of \efswsq. 

This paper describes a measurement of
 \pta ~and \aplfb  ~using the full data sample collected with the OPAL
detector  at LEP during the period 1990-1995
 which corresponds to an integrated luminosity of 151~pb$^{-1}$.
The OPAL measurement of \afb, which is used as input to this analysis,
is described in reference\cite{bib-z0par}.
A sample of \Ntpair
~\eett ~candidate events contained within a fiducial acceptance of
$|\cos\theta_{\tau^-}|<0.90$ is used in the analysis.
Most of the selected events (88\%) were recorded with the centre-of-mass
energy (\ecm) at the peak
of the  \Z resonance
 and the remainder, referred to as `off-peak data',
were recorded  at several distinct \ecm ~values within 3~GeV above and
below the peak. These new results supersede the  measurements
reported in Reference~\cite{bib-OPALPL3},
which were based on an analysis of a 1990-1994 OPAL data sample 
that was restricted to the central region of the detector:
 $|\cos\theta_{\tau^-}|<0.68$.

 The OPAL detector, described in detail in reference~\cite{bib-opal},
 consists of a cylindrical magnetic spectrometer  embedded
 in an electromagnetic (ECAL) calorimeter with presampler
 and a hadronic (HCAL) calorimeter which in turn are enclosed by
 muon detectors (MUON).
 A silicon micro-vertex detector,
  vertex chamber, large volume jet chamber, and z-chambers 
comprise the central tracking detector (CT). These are contained in a 0.435~T
solenoid with magnetic field aligned along the beam axis which
defines the z-axis of the detector.
The ECAL, consisting of a barrel and two endcap arrays of lead glass blocks,
provides complete azimuthal coverage  within a polar angle range of
$|\cos\theta|<0.984$.
The HCAL and muon detectors are collectively referred to as 
the `outer detectors'. 
The detector covers nearly the entire solid angle and provides full
trigger efficiency\cite{bib-trigger}
 for \eett ~events within the acceptance of this analysis.

The  $\tel$, $\tmu$, $\tpi$, $\tK$, $\ttwopi$, $\tKpiz$ and $\threepi$ decays,
representing a combined branching fraction of 83\%,
are selected and their kinematic properties used to measure the polarization. 
As nothing in the selection discriminates between charged pions
and kaons, the $\tpi$ and $\tK$ modes are analysed as a single channel,
denoted as $\tpi$. The  $\ttwopi$ and $\tKpiz$ modes 
are  dominated by the $\rho^{\pm} \ra \pi^{\pm} \pi^0$ 
and K$^{\ast \pm} \ra$K$^{\pm} \pi^0$ resonances, respectively,
and are similarly treated as a combined channel denoted as $\tro$.
As the $\threepi$ decays are dominated by the three-prong 
\footnote{The term `prong' refers to the
 track of a charged particle originating
from the $\tau$ decay vertex.} a$_1$ resonance, this channel is labelled as $\taone$.
The $\tau\ra$K$\pi \pi \nu_{\tau}$ and $\tau\ra$K K$\pi \nu_{\tau}$ 
decay modes are treated as background to the $\threepi$ channel.
The selection criteria for all channels are based on a
likelihood technique and have been optimized 
to reduce the combined statistical and systematic errors.

In order to perform the polarization measurement, the data are fit with
linear combinations of left-handed and right-handed Monte Carlo simulated
kinematic spectra for each of the five channels. 
The five decay modes  do not all have the same sensitivity to the 
$\tau$ polarization. For the simplest
case, the two-body decay of a $\tau$ lepton to a spin-zero $\pi$ meson
 and $\tau$ neutrino,
$\tpi$, the maximum sensitivity is provided by the energy spectrum of
the $\pi$. The pure V-A charged current decay of the $\tau$ together
with angular momentum conservation produces a $\pi$
 momentum preferentially aligned with the
helicity of the $\tau$. In the lab frame, this means that a $\pi^-$
produced from the
decay of a right-handed $\tau^-$ will, on average, be more energetic than
a $\pi^-$ produced from the decay of a left-handed $\tau^-$.
For the leptonic channels, $\tel$ and $\tmu$,
 the additional unobserved neutrino causes a substantial
reduction in polarization sensitivity and the
decay of a right-handed $\tau^-$  produces, on average, a lower energy
charged daughter than a left-handed $\tau^-$.
The $\tro$ channel is complicated by the fact that the spin-1 $\rho$
may be produced in two different spin states, the
third state being forbidden by angular momentum conservation.
The longitudinal state dominates the decay width and is kinematically 
equivalent to the $\pi$ channel. In addition there is a significant
contribution from the allowed transverse state which is also sensitive
to the polarization but in a manner which reduces the
polarization sensitivity if only the total visible energy is measured.
Much sensitivity is regained by spin-analyzing the $\rho^-\ra\pi^-\pi^0$
 decay through measurements
of the kinematics of  the final-state $\pi^-$ and $\pi^0$\cite{DAVIER}.
The $\taone$ channel, which also involves a spin-1 hadron,
 is similar to the $\rho$ channel except that the $\aone$ decays to three
pions. As with the $\tro$ channel, 
sensitivity is optimized by spin-analyzing the $\aone\ra\pi\pi\pi$ decay
via measurements of angles between and momenta of the final state pions.
The details of the specific variables used for each channel are
described Section~\ref{GlobalFittingMethod}.

The maximum sensitivity for each decay mode,
defined as 1/$\sqrt{N}\sigma$ where $\sigma$ is the
statistical error on the polarization measurement using $N$ events,
is given in Table~\ref{table-channelsens} for \ptau=0.0, assuming
that all the information in the decay, apart from the $\tau$ direction,
is used with full efficiency.
A measure of the weight with which a given decay mode ideally
contributes to the overall measurement of the polarization is given
by that decay mode's sensitivity squared multiplied by its branching
ratio. Normalized ideal weights for each decay mode,
 which are calculated assuming
maximum sensitivity  and perfect identification efficiency,
are also given in Table~\ref{table-channelsens}.
As can be seen, the $\tro$ and $\tpi$  channels are expected to dominate the
combined polarization measurement. The actual sensitivity achieved
in the experiment for the selected event sample 
is degraded mainly because of inefficiencies in the
process of selecting a sample of decays and 
by the presence of background in the sample.

The parameters  \pta ~and \aplfb ~are extracted from the data using a global
maximum likelihood fit where the data are described by
linear combinations of positive and negative helicity
distributions in observables appropriate to each $\tau$ decay channel
and in  $\theta_{\tau^-}$. These distributions are
 obtained from  a Monte Carlo simulation in which the $\tau$ lepton
is assumed to decay according to V-A theory. For those
events in which both $\tau$ decays have been classified, the analysis
explicitly takes into account the $\tp$--$\tm$ longitudinal spin correlation
by analysing the event as a whole.
 In so doing, this also accounts for experimental correlations
between the  polarization observables introduced by the
\tn-pair selection and decay mode identification criteria.
A beneficial aspect of performing a global fit to all decay modes
simultaneously is the automatic inclusion
of the correlations between the systematic errors of the different decay
modes.

\begin {table}[htbp]
\begin{center}
\begin{tabular}{|l|c|c|c|c|c|} \hline
                        & \tel    & \tmu    &  \tpi   &\tro     & \taone    \\
  &        &       &      &       &
  a$_1^{\pm}\ra\pi^{\pm}\pip\pim$\\ \hline
 Branching ratio        & \brel   & \brmu   &  \brpi &\brrho   & \braone   \\
 Maximum sensitivity    & \elsensi& \musensi&  \pisensi&\rhosensi& \aonesensi\\
 Normalized ideal weight& 0.06    & 0.06    &  0.30    &0.44     & 0.13      \\
\hline 
\end {tabular}
\caption{The branching ratios, maximum sensitivity and 
  normalized ideal weight for the five decay modes used in the
  analysis. 
  The ideal weight is calculated as the product
  of the branching ratio and the square of the maximum sensitivity.
  Presented in the last line of the table is the ideal weight for each
  channel divided by the sum of the ideal weights of the five channels.}
\label{table-channelsens}
\end{center}
\end{table}

\vspace{12pt}
\noindent
\section{Tau Pair Sample}
The selection of the particular decays used for the measurement can
be viewed as a two stage process where in
the first stage  a sample of \eett candidates is selected from which
samples of
 $\tel$, $\tmu$,  $\tpi$,  $\tro$ and the three-prong $\taone$ decays
 are identified in the second stage.
The $\tau$-pair  selection employs  the same basic criteria used in our 
earlier publications for the barrel region of the detector
~\cite{bib-OPALPL3,bib-OPALPL2, bib-OPALPL1} with slight modifications 
to allow for an extension of the acceptance into the endcap regions. These
are described in detail, along with all aspects of this analysis,
in Reference~\cite{bib-graham}.
 The general
strategy is to identify events characterized by a pair of
back-to-back, narrow jets with low particle multiplicity (\tjet). 
Background from two-photon processes is then suppressed
by requiring that the events have a minimum total visible energy and
significant missing transverse momentum when the total energy
in the event is low. After removing cosmic ray backgrounds,
the events which remain are almost entirely lepton-pairs.
Events with high measured energy that are consistent
 with being \eeee or \eemm are also removed.
 The polar angle of each \tjet ~with respect to the direction of the 
 e$^-$ beam, $\theta_{jet}$, is determined using
 charged tracks and clusters of deposited energy in the  ECAL.
 Events are selected if the average of  $\act$ for the two \tjets ,
 $\overline{\act}$, is less than 0.90. The same  $\overline{\act}$
 is used as the estimator for the magnitude of  $\cos\theta_{\tau^-}$
 in the analysis. 
 Using this selection, a sample of  \Ntpair ~events is obtained.

The contributions to the selected events from 
various physics processes are estimated 
 using a number of Monte Carlo data samples which have approximately ten
times the number of events as the data. The \eett signal and
\eemm background are both modelled using the {\mbox KORALZ} Monte Carlo 
generator with the TAUOLA decay package~\cite{bib-koralz} and 
 the \eeee background is estimated using the
{\mbox BHWIDE} generator~\cite{bib-bhwide}. The residual 
$\ee\ra\qq$ hadronic
background is simulated using the {\mbox JETSET} Monte Carlo~\cite{bib-Jetset}
 with parameters
tuned to fit the global event shape distributions of OPAL 
data~\cite{bib-OPALQCD}. Contributions from 
non-resonant t-channel two-photon processes (\eeeeee, \eeeemm )
 are estimated using the
generator described in Reference~\cite{bib-vermas} whilst hadronic
two-photon processes are modelled using the generator documented in 
Reference~\cite{bib-hadronictwophoton}.
The potential effects of four-fermion events, including those with hadronic 
final states, are studied using 
the generator described in Reference~\cite{bib-FERMISV}.
The response of the OPAL detector to the generated particles in each
case is modelled using a simulation program~\cite{bib-gopal} based on
the GEANT~\cite{bib-geant} package. In all cases, the Monte Carlo and
real data are treated  in an identical manner.
Estimates of the \eett selection efficiencies and purities within the
different fiducial regions of the analysis are obtained 
using these Monte Carlo samples and are presented in Table~\ref{table-effpure}.

\begin {table}[htbp]
\begin{center}
 \begin{tabular}{|c|c|c|c|}
 \hline
 Region & $\cos{\theta}$     & Efficiency & Purity\\
 \hline
Barrel & $\cos{\theta} \le 0.72$      &    93 \%   &  98 \%\\
Overlap & $0.72<\cos{\theta} \le 0.81$ &    75 \%   &  98 \%\\
Endcap &$0.81<\cos{\theta} \le 0.90$  &    79 \%   &  97 \%\\
\hline 
\end {tabular}
\caption{$\tau$-pair selection  efficiencies 
and purities for the different fiducial regions of the analysis.
}
\label{table-effpure}
\end{center}
\end{table}

\section{Tau Decay Selection}
Starting with this sample of $\tau$-pair candidates, each \tjet ~in an event is
classified as one, and only one, of \tel , \tmu , \tpi , \tro  ~or \threepi ,
or none of these, using a likelihood-based identification procedure.
Any event having at least one identified \tjet ~is then used in
 the polarization analysis. From this procedure \NtpairSel ~events
 contribute to the measurements  out of the first-stage selection
 of \Ntpair ~events.

Each of the five decay mode selections start with a mode-dependent
set of loose cuts which define a sample subsequently processed
in a binned likelihood selection\cite{bib-Karlen}.
 The likelihood selection procedure uses observables
which provide discrimination between various decay channels of the 
 $\tau$ lepton and non-$\tau$ background. The Monte Carlo simulation provides 
normalized distributions for a set of observables, $O_i$, for each of the
decay modes. These are subsequently used to calculate for each decay 
channel $j$,
the likelihood, $\ell^j_i(O_i)$,  that the measured $O_i$ would be observed.
The likelihood that decay mode $j$ produces
the measured observables in a given \tjet ~is obtained from
the product of the likelihoods: ${\cal L}(j) = \prod_i \ell^j_i(O_i)$.
In order to select decays from mode $k$, a cut is applied to its relative
likelihood, $L(k) = {\cal L}(k) / \sum_j {\cal L}(j)$. 
From this definition, $L(k)$ lies between 0 and 1 and the value of the
cut is chosen to maximize the product of purity and efficiency. The
requirement that decays  have large values of $L(k)$ produces a sample with low
background normally at the cost of efficiency for selecting mode $k$ decays.
In this work, each of the five decay mode selections employ different
observables and therefore exploit a different set of likelihoods,
${\cal L}(j)$. If a \tjet ~is classified in more than one channel after
applying the cut
to the likelihood, then it is reclassified into the channel having the
largest relative likelihood.

Before applying the likelihood selection, fiducial requirements are 
imposed to remove  the small fraction of decays having
particles entering regions of the detector which are inadequately modelled
by the Monte Carlo simulation. 
Furthermore, the discrimination between channels is enhanced by
 dividing the preseleced
sample into a set of subsamples according to the detector $\act$ region
(barrel, overlap or endcap as defined in Table~\ref{table-effpure})
and other \tjet ~characteristics, such as the number of electromagnetic
clusters in the jet unassociated with tracks.

Approximately 25\% of $\tau$ leptons decay to $\nu_{\tau}$ and a \rhon 
~meson, which subsequently decays almost exclusively to a charged 
and neutral pion. Consequently, 
much of the discrimination between the $\tro$ and the other
decay modes of the $\tau$ is  achieved by identifying
 photons from a single $\pi^0$ decay which form an invariant mass
 with a single charged track consistent with the $\rho$ mass, m$_{\rho}$.
 The ECAL observables use a `maximum entropy'
clustering algorithm\cite{bib-maxent} which is well suited to identifying
photons in a \tjet. A cluster
in the barrel (overlap, endcap) region of the detector
 which is not associated with 
a charged track is referred to as a `neutral cluster' if it  has
an energy of at least 650~MeV (1.25~GeV, 1.0~GeV).
 When there is only one neutral cluster present in the \tjet
 ~then it is considered as a $\pi^0$ candidate.
 If there are at least two neutral clusters 
then the invariant mass of the two most energetic
clusters forms an observable, m$_{1,2}$, and the invariant 
mass of this object with the charged track forms the observable
referred to as m$_{\rho}$. A third invariant mass, m$_{jet}$, is 
reconstructed from all neutral clusters and charged tracks in the \tjet ~and
a fourth, m$_{charged}$, is reconstucted from all charged tracks in the \tjet.
Since 9\% of $\tau$ leptons decay via \tpitwopiz, and these can appear as
 a background in the $\tro$ channel,
 a fifth invariant mass, m$_{1-prong}$, is reconstructed from
the highest momentum track and the neutral clusters in the \tjet. If there
are more than four neutral clusters, then only the four highest energy
clusters are used.
These various invariant masses are used as likelihood variables.

The other observables exploited in the likelihood selections are:
information from the specific energy loss of the charged track
as measured in the central jet chamber (dE/dx); 
the azimuthal angle between the highest momentum charged track and the 
presampler cluster closest to that track ($\phi_{pres}$);
the number of neutral clusters;
the ECAL energy associated with the 
highest momentum track in the \tjet ~(E$_{ass}$);
the ratio of $E_{ass}$ to the momentum of the highest momentum track
in the \tjet ~(E$_{ass}$/p); 
 the ratio of the ECAL energy of the highest energy cluster 
 to the momentum of the track with the highest momentum 
 in the \tjet ~(E$_{max}$/p); 
the ratio of the  ECAL energy measured in the \tjet ~from neutral clusters
and clusters associated to tracks 
to the momentum of the track having the 
highest momentum in the jet (E$_{jet}$/p); 
 the energy of the neutral  clusters not used in the calculation
 of m$_{\rho}$ (E$_{resid}$); 
the energy of the neutral clusters used to calculate m$_{1,2}$ (E$_{1,2}$);
 a variable describing how well the track
from the central detector matches a track segment in the muon detectors
(CT-MUON);
 and information from hits measured in  HCAL and MUON.

\begin {table} [htb]
\begin{center}
 \begin{tabular}{|l|ccccc|} \hline
  Observable             & e & $\mu$&$\pi$& $\rho$ & $\aone$ \\ \hline
  dE/dx                  & X &   X  &  X  &    X   &    X     \\
 $\phi_{pres}$           &   &   X  &  X  &        &          \\
 No. neutral clusters    &   &      &  X  &    X   &          \\
 E$_{ass}$               &   &   X  &     &        &          \\
 E$_{ass}$/p             & X &   X  &  X  &        &    X     \\
 E$_{max}$/p             &   &      &     &        &    X     \\
 E$_{jet}$/p             & X &      &  X  &    X   &    X     \\
 E$_{resid}$             &   &      &     &    X   &          \\
 E$_{1,2}$               &   &      &     &    X   &    X     \\
 m$_{1,2}$               &   &      &  X  &    X   &          \\
 m$_{\rho}$              & X &   X  &  X  &    X   &          \\
 m$_{jet}$               &   &      &  X  &    X   &          \\
 m$_{charged}$           &   &      &     &        &    X     \\
 m$_{1-prong}$           &   &      &     &    X   &    X     \\
 CT-MUON 		 & X &   X  &  X  &    X   &    X     \\
 HCAL hits               &   &   X  &  X  &    X   &    X     \\
 MUON hits               &   &   X  &  X  &    X   &    X     
\\ \hline
\end{tabular}
\caption[y]{Observables employed in the likelihood selections used to 
classify the different decay modes. An `X' indicates that the observable
is used in forming the likelihood distribution for the indicated
decay mode selection.}
\label{table-Likevars}
\end{center}
\end{table}

\begin{figure} 
  \begin{center}
  \mbox{\epsfig{file=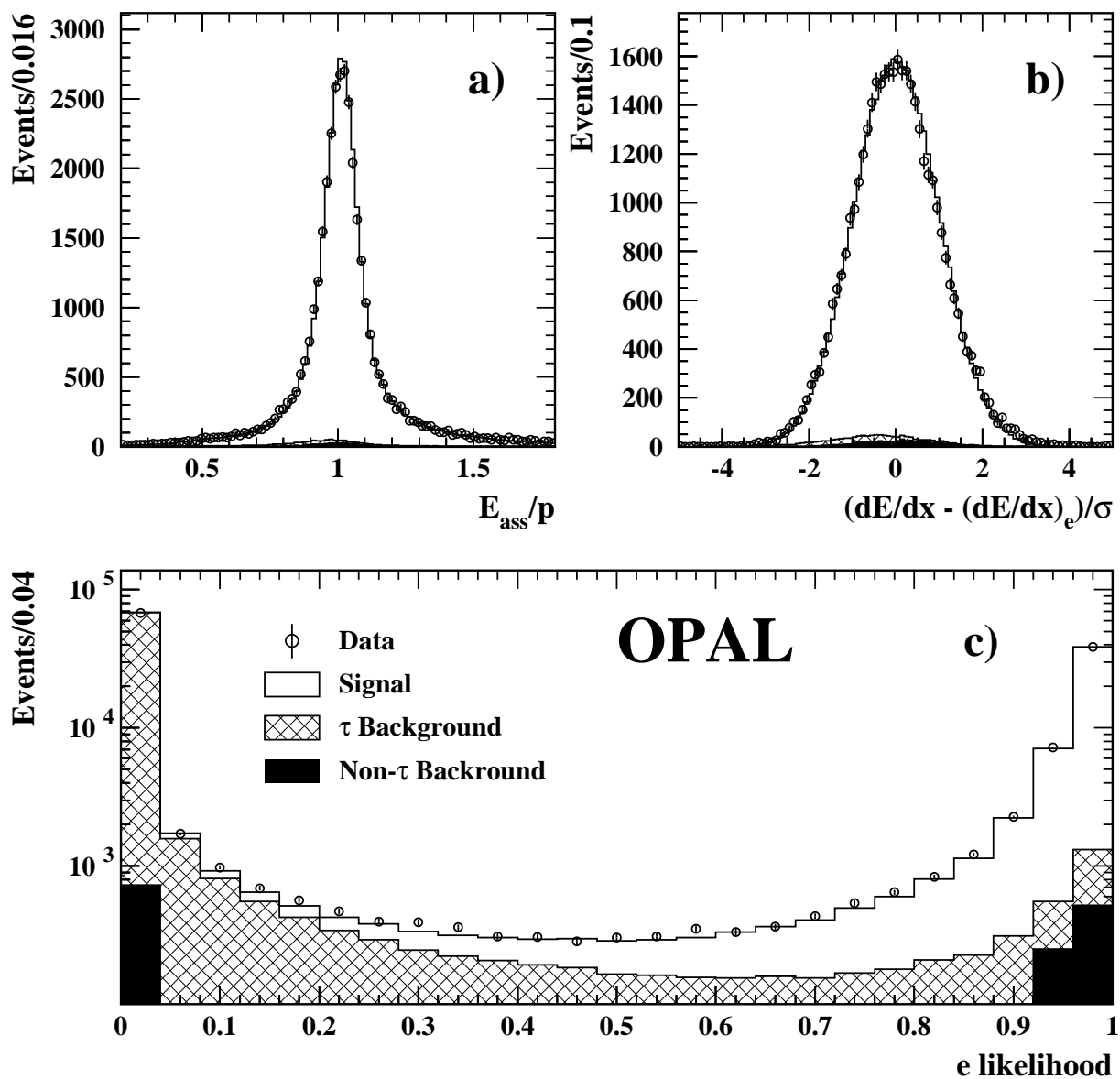,width=18cm,height=18cm}}
 \end{center}
\caption[]{\small
 Distributions of (a) E$_{ass}$/p and
(b)  pull distribution of measured dE/dx
compared to the dE/dx expected under an electron hypothesis
for  selected  \tel candidates.
(c) The distribution of likelihoods formed from all observables used
in the \tel ~selection. 
In each figure  the points with error bars represent the data, the open
 histogram  the
 \tel 
~expectation from Monte Carlo, the hatched histogram
the  cross-contamination from other $\tau$ decays and the
dark shaded histogram the background from non-$\tau$ sources.
The distributions include data from all detector regions and subsamples.
}
{ \label{fig-electron} }
\end{figure}

\begin{figure} 
  \begin{center}
  \mbox{\epsfig{file=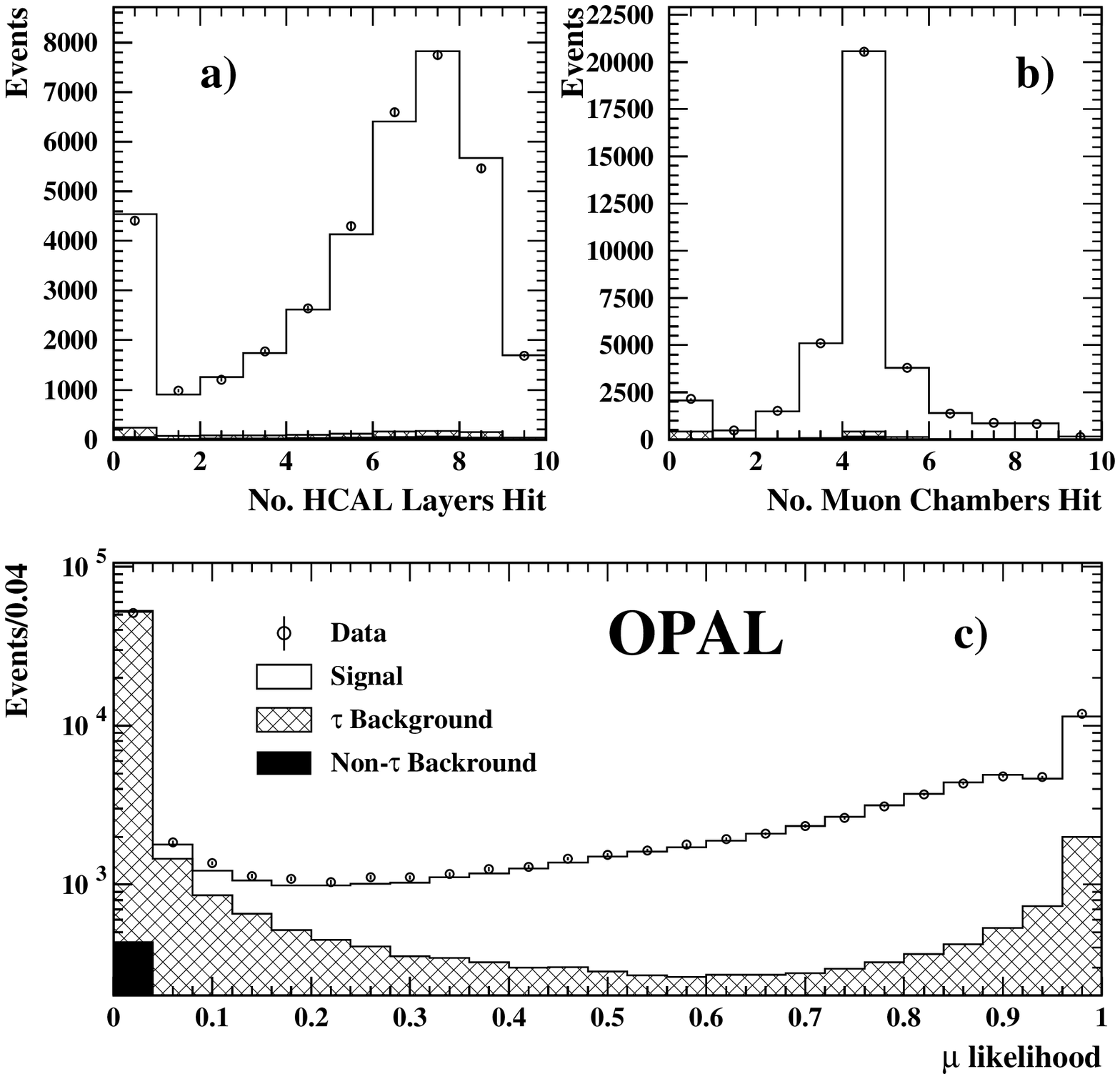,width=18cm,height=18cm}}
 \end{center}
\caption[]{\small
 Distributions of (a) the number of HCAL layers hit and
 (b)  the number of muon chambers hit in the \tmu ~selected jets.
(c) The distribution of likelihoods formed from all observables used
in the \tmu ~selection.
In each figure  the points with error bars represent the data, the open
 histogram  the
 \tmu 
~expectation from Monte Carlo, the hatched histogram
the  cross-contamination from other $\tau$ decays and the
dark shaded histogram the background from non-$\tau$ sources.
The distributions include data from all detector regions and subsamples.
}
{ \label{fig-muon} }
\end{figure}

The following subsections provide more detail about the likelihood selections
for each of the five channels. In particular, the subsections
specify for each decay  mode: the preselection cuts; the
subsamples upon which likelihood selections are separately applied
within the three $\cos\theta$ regions of the detector;
the observables employed in the likelihood selections,
which are summarized in Table~\ref{table-Likevars};
efficiencies within the fiducial acceptance after tau pair selection;
the amount of cross-contamination
from other $\tau$ decay modes; and the level of background
 from non-$\tau$ sources. 
 The observables exploited in the likelihood selections are studied and
in general are well described by the Monte Carlo simulation.
Any differences between the data and simulation of these observables
are taken into account in 
the assessment of the systematic uncertainties on the polarization
measurements. Distributions of two observables characterising each
decay mode are plotted in Figures~\ref{fig-electron}(a,b)-\ref{fig-aone}(a,b)
as representative indications of the level of agreement between data
and the Monte Carlo simulation.
The distributions of the likelihoods formed from all observables used in each
of the five selections are plotted in 
Figures~\ref{fig-electron}(c)-\ref{fig-aone}(c). Each of these plots comprise a
sum of the various likelihood distributions used for each decay mode selection
and further illustrate the adequacy of the detector modelling.
The likelihood cut values are optimized separately
for each of the various likelihood distributions,
 and not on the combined distributions plotted here.

\subsection{$\boldmath{\tn \ra}$e$\boldmath{\nueb \nut}$ ~identification}
The preselection of the $\tel$ sample requires, within the \tjet,
~that there be no more than three neutral clusters; that
no more than three of nine HCAL layers register activity; that
no more than three  muon chambers register activity; and 
that there be one or two tracks. If there are two tracks, it is
assumed that the higher momentum track is associated with the electron
from the $\tau$ decay and the lower  momentum track is ignored.
The observables used to form the likelihoods include: 
 dE/dx, E$_{ass}$/p, E$_{jet}$/p, CT-MUON,
and m$_{\rho}$. The likelihoods in the different
regions of the detector are formed for a subsample of \tjets ~with
the number of neutral clusters equal to zero and for a sample
with at least one neutral cluster.
 Distributions of two of the
variables used in this selection, E$_{ass}$/p, and the pull of the measured
dE/dx under an electron hypothesis are shown in 
Figure~\ref{fig-electron}. The final selection produces a 
sample of  \Nel ~candidates 
with an efficiency of \eleff ~after tau pair selection
 within the fiducial region and a
background  of \elbkg.
Most of the cross-contamination from $\tau$ decays 
 arises from  \tpi ~decays (0.9\%)
and from \tro ~decays (1.3\%) along with contributions from a number of other
channels (0.8\%). 
The non-$\tau$ background is estimated to contribute approximately 1.6\%.

\subsection{$\boldmath{\tmu}$ ~identification}
The likelihood selection for the  \tmu ~decays is applied to those
\tjets ~preselected to have a single charged track, no more than two
neutral clusters and a measured track momentum greater than 4\% of the
beam energy.  This sample is divided into subsamples based on 
information from the outer detectors before forming the likelihoods for
the different detector regions.
The observables used in this selection include: 
dE/dx, E$_{ass}$, E$_{ass}$/p, $\phi_{pres}$, CT-MUON,
 information from the HCAL and muon chambers,
 and m$_{\rho}$.  The distributions of the number of HCAL
layers hit in the \tjet ~and the number of muon chambers 
hit for all selected \tmu ~decays for both data and Monte Carlo
simulation of signal and background are displayed in Figure~\ref{fig-muon}.
This selection has an efficiency of \mueff ~after tau pair selection
within the fiducial acceptance
and background of \mubkg ~which is heavily dominated by
 the \tpi ~decays with some components arising 
 from the \eemm~and two-photon processes. This results in the
 selection of  \Nmu ~decays.

\begin{figure} 
  \begin{center}
  \mbox{\epsfig{file=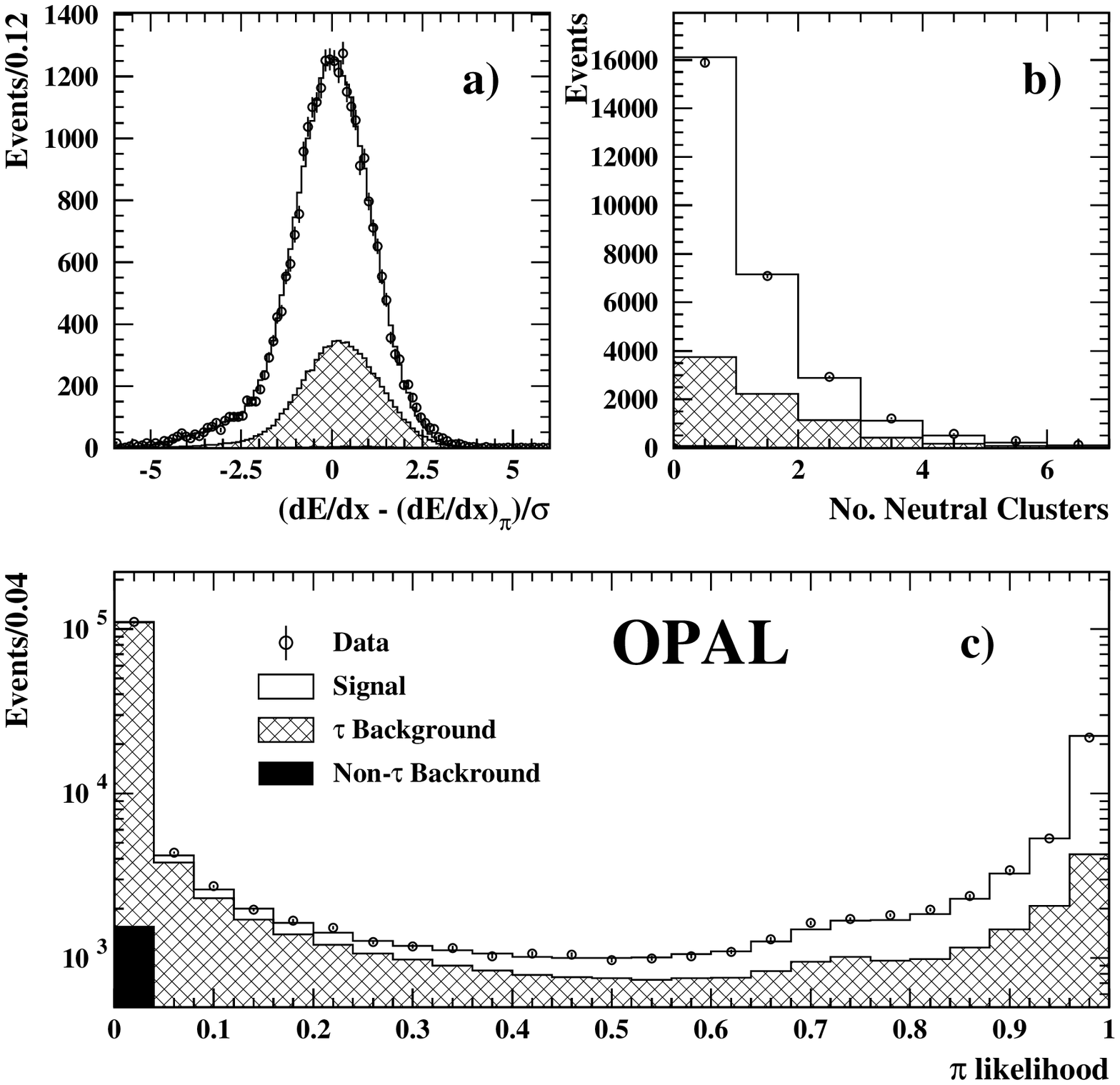,width=18cm,height=18cm}}
 \end{center}
\caption[]{\small
 Distributions of (a)  pull distribution of measured dE/dx
compared to the dE/dx expected under a pion hypothesis and (b)
the number of neutral clusters.
(c) The distribution of likelihoods formed from all observables used
in the \tpi ~selection. 
In each figure  the points with error bars represent the data, the open
 histogram  the
 \tpi
~expectation from Monte Carlo, the hatched histogram
the  cross-contamination from other $\tau$ decays and the
dark shaded histogram the background from non-$\tau$ sources.
The distributions include data from all detector regions and subsamples.
}
{ \label{fig-pion} }
\end{figure}

\subsection{$\boldmath{\tpi}$ ~identification}
The  \tpi ~decay preselection requires that the \tjet~
contains one track and that the ratio of the momentum of that track to
 the beam energy be at least 0.02. Within the different detector regions
this sample is divided into subsamples with zero, one or more than
one neutral cluster, and subsamples where at least one HCAL layer records
 hits or where no HCAL layers record hits. For these various samples
likelihoods are formed from the following observables: 
dE/dx, E$_{ass}$/p, E$_{jet}$/p,  $\phi_{pres}$,
the number of neutral clusters, CT-MUON,
information from the HCAL
and muon chambers, m$_{1,2}$, m$_{\rho}$, and m$_{jet}$.
Representative distributions of two of the variables important in this
analysis are seen in Figure~\ref{fig-pion} for those \tjets ~selected
as \tpi ~candidates:  the pull of the measured
dE/dx under a pion hypothesis and the number of 
neutral clusters.
A total of \Npi ~\tpi ~candidates are selected with an efficiency of
 \pieff ~after tau pair selection 
within the fiducial region with a background level of \pibkg.
Most of the background arises from the \tro ~mode (16\%) with the
next largest contributions arising from \tmu ~(5\%) and \taone ~(2\%).
The non-$\tau$ background is estimated to contribute
approximately 0.2\%.

\begin{figure} 
  \begin{center}
  \mbox{\epsfig{file=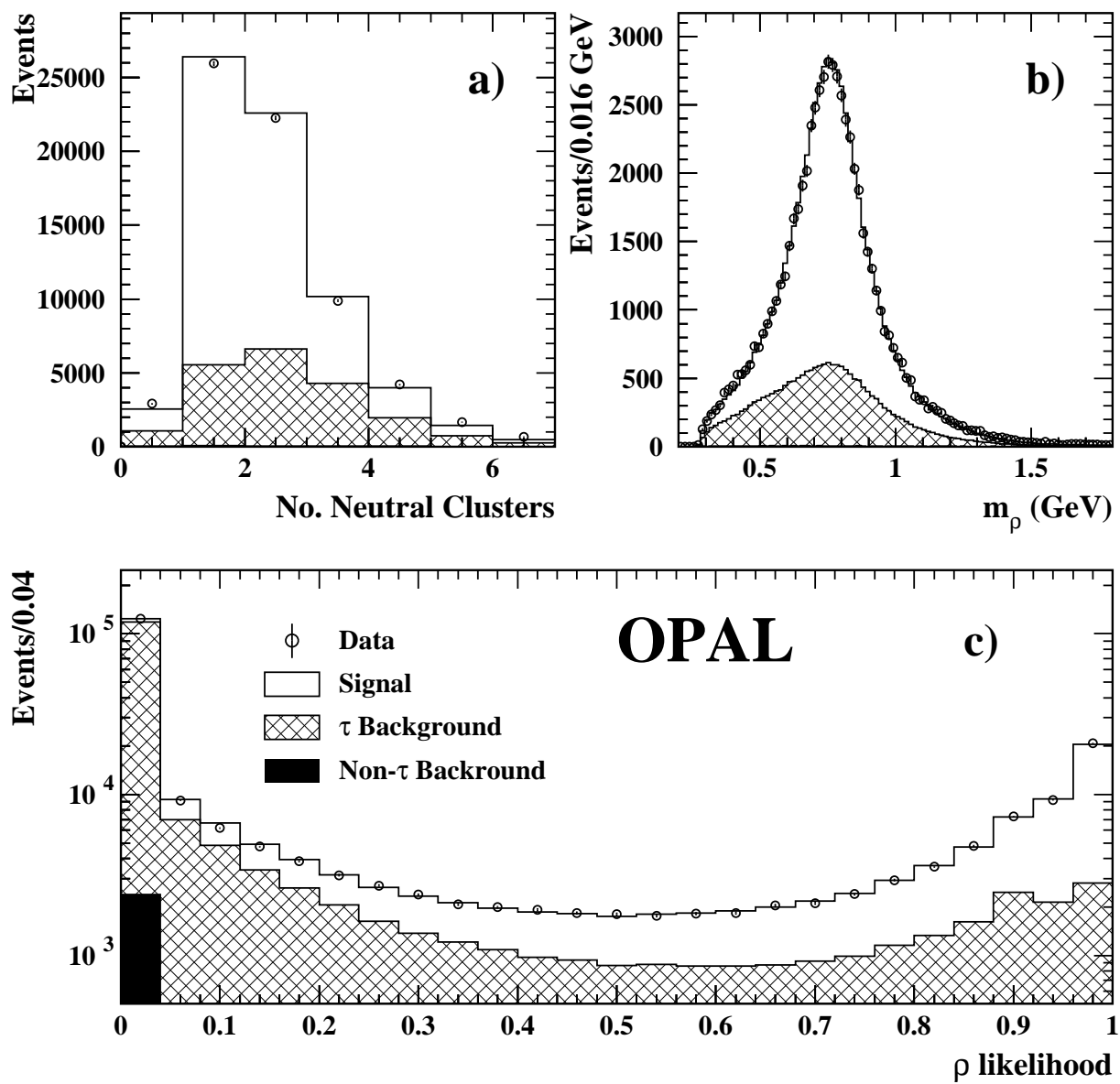,width=18cm,height=18cm}}
 \end{center}
\caption[]{\small
 Distributions of 
 (a) the number of neutral clusters for \tro ~candidates and
(b) the reconstructed $\rho$ mass.
(c) The distribution of likelihoods formed from all observables used
in the \tro ~selection. 
In each figure  the points with error bars represent the data, the open
 histogram  the
 \tro
~expectation from Monte Carlo, the hatched histogram
the  cross-contamination from other $\tau$ decays and the
dark shaded histogram the background from non-$\tau$ sources.
The distributions include data from all detector regions and subsamples.
}
{ \label{fig-rho} }
\end{figure}

\subsection{$\boldmath{\tro}$ ~identification}
The  \tro ~decay preselection consists solely of
 the requirement that the \tjet~
contains one track.
This sample is divided into the subsamples with
  zero, one or more than
one neutral cluster and subsamples where at least one HCAL layer records
 hits and where no HCAL layers record hits. A number of the observables 
used to create the likelihood selection are  also used
for the \tpi ~likelihood: 
dE/dx, the number of neutral clusters,
E$_{jet}$/p, CT-MUON, 
 information from the HCAL and muon chambers,
 m$_{1,2}$,  m$_{\rho}$, and m$_{jet}$. Additional
observables used to enhance the \tro ~selection include:
E$_{resid}$, E$_{1,2}$, and m$_{1-prong}$.
The distributions of two of the significant observables used in
this selection are displayed in Figure~\ref{fig-rho} for those \tjets ~selected
as \tro ~candidates: m$_{\rho}$ and the number of 
neutral clusters. 
The analysis selects \Nrho ~\tro ~candidates from the $\tau$-pair 
sample. Within the polar-angle acceptance, the efficiency is
\rhoeff. The background fraction in the \tro ~sample is \rhobkg ~and consists
 mainly of \thnpiz ~(18\%) and \tpi ~(5\%) decays.
 The non-$\tau$ background is estimated to contribute
 0.2\%.

\begin{figure} 
  \begin{center}
  \mbox{\epsfig{file=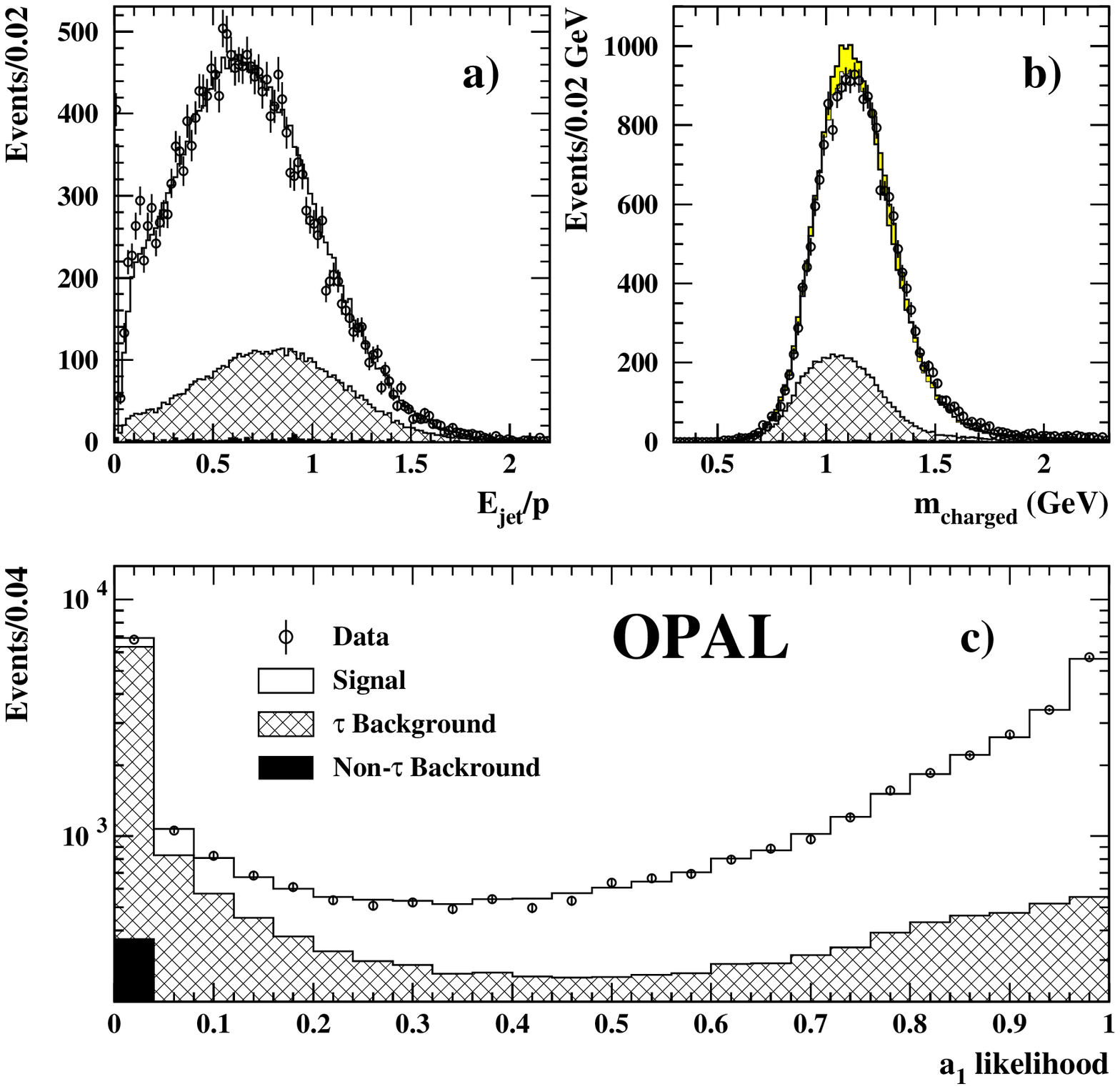,width=18cm,height=18cm}}
 \end{center}
\caption[]{\small
 Distributions of
 (a) the ratio of the total ECAL energy in the \tjet ~to
the momentum of the track with the highest momentum in the \tjet ~and
 (b) the reconstructed \aone ~mass for \taone ~candidates.
(c) The distribution of likelihoods formed from all observables used
in the \taone ~selection. 
In each figure  the points with error bars represent the data, the open
 histogram  the
 \taone
~expectation from Monte Carlo, the hatched histogram
the  cross-contamination from other $\tau$ decays and the
dark shaded histogram the background from non-$\tau$ sources.
The lightly shaded region of the  histogram in (b) represents
the \taone ~decay modelling uncertainty.
The distributions include data from all detector regions and subsamples.
}
{ \label{fig-aone} }
\end{figure}

\subsection{$\boldmath{\taone}$ ~identification}
For the \taone ~selection we restrict ourselves to the three-prong mode,
which has a branching fraction of 9\%. 
 It is assumed that all three-pion decays
of the $\tau$ lepton proceed through the \aone ~channel\cite{bib-a1_3prong}.
 The \taone ~\tjet ~is required to
have three charged tracks, none of which is identified
 as a conversion electron.
The major contamination for this mode is from
 $\tau \ra 3\pi\geq1\pi^0\nu_{\tau}$
 decays.
Another potential source of background arises from \eeee ~events containing a
conversion, therefore some of the likelihood variables included 
for the \taone ~selection are designed to suppress this contamination.
The likelihoods are formed from: dE/dx,
E$_{ass}$/p, 
E$_{jet}$/p, 
E$_{max}$/p,
E$_{1,2}$,
CT-MUON,
information from the HCAL and muon chambers,
 m$_{1-prong}$, and m$_{charged}$. 
The distributions of E$_{jet}$/p and
 m$_{charged}$ are plotted in Figure~\ref{fig-aone}
 for those \tjets ~selected as \taone ~candidates for both
the data and Monte Carlo simulation of signal and background.
The differences between the data and Monte Carlo simulation evident
in these plots are adequately described by the systematic uncertainties 
assessed for this analysis.  In particular, the differences between
the data and Monte Carlo
distributions of m$_{charged}$, Figure~\ref{fig-aone}(b), 
are well described by the systematic 
uncertainty ascribed to the modelling of the $\taone$ decay from the use of 
different $\aone$ models\cite{bib-Kuhn,bib-IMR}. This modelling
uncertainty is shown by
the lightly shaded region of the  histogram in Figure~\ref{fig-aone}(b).
The number of selected \taone ~candidates is \Naone. 
The selection efficiency is  \aoneeff ~and the  background is \aonebkg,
~most of which is from $\tau \ra 3\pi\geq1\pi^0\nu_{\tau}$ 
and \Kpipi ~decays (20\%) and some of which contains residual conversions
from \tro ~events. The non-tau background in the sample
is estimated to be 0.2\%.

\subsection{Survey of Pair-identification Classes}
\label{subsec-topo}
The fitting approach adopted in this analysis lends itself to a 
classification of the data in which events are grouped according to the
identified decay mode of each of the two $\tau$ decays in the event.
For example, if in a given event one $\tau$ decay is identified
 as \tel ~and the other as \tpi
~then this event belongs to one pair-identification class,
 which is denoted as `$\pi$~e'.
There are 20 such statistically independent
classifications including those in which one of 
the $\tau$ decays is identified and the other is not. An example of the
latter would be the pair-identification class
denoted as `$\rho$~nid' in which one of the $\tau$ decays is identified
as \tro ~and the other is not identified.
The numbers of data events for each of the 20 possible event
 pair-identification
classes are listed in Table~\ref{table-topoevents}. Also shown in the
table is the expected number of events from Monte Carlo estimates from $\tau$
and non-$\tau$ sources, scaled by the absolute luminosity. The good
 agreement between the data and the expected numbers of events
in the different pair-identification classes
 evident from this table helps validate the overall
efficiencies and purities estimated for each class.

\begin {table} [htb]
\begin{center}
 \begin{tabular}{|r|c|c|c|c|c|c|} \hline 
    &  nid    &   e  & $\mu$ & $\pi$  & $\rho$  & $\aone$  \\ \hline
 e     &12053   &3599      &       &           &         &          \\
       &11793   &3584      &       &           &         &          \\
\hline
 $\mu$ &11406   &6737      &2746       &           &         &          \\
       &11211   &6601      &2707       &           &         &          \\
\hline
$\pi$ &8382   &4786      &4302       &1669           &         &
\\
       &8200   &4847      &4325       &1645           &         &
\\ \hline
$\rho$ &18006   &10649      &10178       &7504           &8225         &
\\                         
       &18186   &10782      &10230       &7460           &8482         &
\\ \hline
$\aone$&5503   &3736      &3537       &2428           &5631         &867
\\
       &5630   &3753      &3581       &2548           &5669         &915
\\ \hline
\end{tabular}
\caption[y]{Number of  \tn-pair events in each pair-identification class
is presented as the first number in each cell. The expected number of
events from Monte Carlo estimates using absolute luminosity scaling are
shown on the second line.
 The label `nid' refers to the case where the $\tn$ decay was not
 identified.
}
\label{table-topoevents}
\end{center}
\end{table}

\vspace{12pt}
\noindent
\section{Global Fitting Method}
\label{GlobalFittingMethod}
For the measurements of \pta ~and \aplfb,
the distributions described in Equation~\ref{dspmdcst1}
cannot be directly
measured as it is not possible to determine the $\tn$ helicity
on an event-by-event basis. Instead,
distributions of kinematic variables of the $\tn$ decay products
which depend on the \tn ~helicity are used.
These variables, as well as their distributions, depend
on the decay mode analysed\footnote{Note that the
distributions are the same for the \tp ~and \tm ~provided that \pta
 ~is taken as the \tm ~helicity.}. 

In general, for each \tn ~decay channel, $i$, the decay distribution 
depends on a set of kinematic variables, $\vec{\xi_i}$. For
positive helicity states the decay distribution can be expressed as
$f_i(\vec{\xi_i})+ g_i(\vec{\xi_i})$ whilst for negative helicity states
the distribution is $f_i(\vec{\xi_i})- g_i(\vec{\xi_i})$. Consequently,
the measured decay distribution
depends linearly on the weighting of the two helicity states,
 \Pt ~\cite{bib-Tsai}:
\beq
\label{wx}
\frac{1}{\Gamma_i}\frac{d^n \Gamma_i}{d^n \vec{\xi_i} }=
f_i(\vec{\xi_i})+\Pt g_i(\vec{\xi_i}).
\eeq

For  \tel , \tmu  ~and \tpi ~decays, $\vec{\xi_i}$ is one dimensional 
where the relevant kinematic variable ($x_e$, $x_{\mu}$ or $x_{\pi}$)
 is the charged particle
 energy scaled by the beam energy.  For  \tel ~decays, the energy
 measured in the ECAL associated with the \tjet ~($x_e$) is used,
 whereas for  \tmu ~and \tpi ~decays, the energy is
 determined using  the  momentum of the charged particle measured in the
 central tracking detector.

For \tro ~decays, three kinematic variables enter into the polarization
analysis:
$\theta^*$, the angle of the $\rho$ momentum relative to the \tn ~flight
direction  in the \tn ~rest frame;
$\psi$, the angle of the charged pion relative to the $\rho$
flight direction in the $\rho$ rest frame; and m$_{\rho}$, the invariant
 mass of the charged particle under a $\pi^+$ hypothesis and the $\pi^0$.
 This spin-analysis of the $\rho$ decay recuperates
most of the sensitivity which would otherwise be lost if only
the charged pion momentum were used, as discussed in 
Section~\ref{sec-Introduction}.  These three variables
 can be converted into a single optimum
variable, $\omega$, with no polarization sensitivity loss~\cite{DAVIER}.
The variable $\omega$ is defined by
  $\omega$=$g/f$=(R$_{\mathrm +}$ -- R$_{\mathrm -}$ )/(R$_{\mathrm +}$ + R$_{\mathrm -}$)
 where  R$_{\mathrm +}$ and R$_{\mathrm -}$  are the population
densities of positive and negative helicity $\tau$ lepton decays,
 respectively,  which are functions of $\theta^*$, $\psi$ and m$_{\rho}$
\cite{DAVIER}.

The \taone ~channel is more complicated because the
\aone ~decays into three pions. 
Six observables are used in order to improve the
sensitivity in the \taone ~channel\cite{DAVIER}:
the angle   ($\theta ^{*}$)
between the \aone ~and \tn ~momenta in the $\tau$
 rest frame,
the angle ($\psi$) between the perpendicular to the \aone ~decay plane
and the \aone ~flight direction in the rest frame of the \aone ,
the angle  ($\gamma$) in the \aone ~rest frame between the
unlike-sign pion momentum in the \aone ~rest frame and the \aone
~flight direction projected into the \aone ~decay plane,
the $3\pi$-invariant mass, and the two $\pi^+ \pi^-$
 mass combinations present in the
 $\mathrm{a}_1^{\pm}\ra\pi^{\pm}\pi^+\pi^-$ decay.
The distribution of the invariant mass of the three charged particles
assuming them all to be pions, shown in Figure~\ref{fig-aone}b,
demonstrates that agreement between the data
 and simulation of this quantity is reasonable within the uncertainties
of the modelling of the \taone ~decay.
The Monte Carlo distribution depends on the mass and width of the
\aone ~as defined within the framework of a particular model
 of \taone ~decay\cite{bib-Kuhn} and allowance in the assignment of
 systematic errors is made for \taone ~model dependence. As with the 
\tro, these  observables are converted into a single optimum
variable, $\omega$, which this time depends on
the population densities of positive and negative helicity $\tau$
 lepton decays  which are functions of the six variables mentioned above.

The joint distributions of the \tn-pair production and decay can be
expressed in the improved Born approximation as:
\begin{eqnarray}
\label{dsig3}
\frac{d^3\sij}{d\cos\theta_{\tau^-}\:d\xxi\:d\xxj} & = &
  \thovsxt\sij\sum_{\lambda=\pm1} [
             (1 + \cstsq + \eitovth \afb \cst)  + \\
 & & 
  \lambda(\pta (1+\cstsq)+  \eitovth \aplfb \cst) ] \times \nonumber \\
 & &
   [F_i(x_i,|\cos\theta_{\tau^-}|) + \lambda G_i(x_i,|\cos\theta_{\tau^-}|)]
   [F_j(x_j,|\cos\theta_{\tau^-}|) + \lambda G_j(x_j,|\cos\theta_{\tau^-}|)], 
 \nonumber
\end{eqnarray}
where $\sij$ is the cross-section to produce an \eett ~event in which
one \tn ~decays via channel $i$ and the other via channel $j$.
The first two lines of Equation~\ref{dsig3} refer to the production of
the $\tau$-pairs and the third line to the $\tau$ decays.
The summation over $\lambda$ indicates that the summation is over 
positive and negative helicities. 
The symbol $x_i$ represents the kinematic  variable corresponding
to channel $i$:
 $x_{\eln}$, $x_{\mun}$, $x_{\pin}$, $\omega_{\rhon}$ or $\omega_{\aone}$.
 The decay distributions 
 for positive-helicity $\tau$ leptons are given by
 $F_i+G_i$ whereas the  decay distributions for negative-helicity $\tau$ 
leptons are given by  $F_i-G_i$. $F_i$ and $G_i$ represent
functions of $x_i$ and  $|\cos\theta_{\tau^-}|$
after including the effects of the decay mode identification procedure,
 detector response and radiation. 
  The simulation of the underlying detector measurements that go into 
 each of the observables used in the analysis, such as track momentum
and ECAL cluster energies and positions, is checked
 and corrected if necessary, using various control samples as discussed below.
Equation~\ref{dsig3} includes the correlation between the decay
distributions of the two \tn ~leptons
when analysing events in which both \tn ~decay
channels are identified. 

 A binned maximum likelihood fit is performed to simultaneously extract
 \pta ~and \aplfb  ~by fitting the linear combination of
the positive and negative helicity Monte Carlo distributions to the
data. The values of $x_i$, $x_j$ and $\cos\theta_{\tau^-}$ for each
event are  calculated and a histogram, binned\footnote{
There are ten bins in $\cos\theta_{\tau^-}$ and ten bins in each of
 $x_{\eln}$, $x_{\mun}$, $x_{\pin}$, $\omega_{\rhon}$, and $\omega_{\aone}$.}
 in $x_i$, $x_j$ and
$\cos\theta_{\tau^-}$, is then filled for each \ecm.
A value for   \cst ~of the event is determined from $\overline{\act}$
and the sign of the charge of the identified $\tau$ decay.
A separate set of  histograms exists for each combination of decay channel
 pairs. If only one $\tau$ decay
is identified, then only bins in  $x_i$ and $\cos\theta_{\tau^-}$ are filled.
The same procedure is performed for the Monte Carlo with a
separate set of histograms filled for the positive and negative helicity
$\tau$ leptons binned in  $x_i$, $x_j$ and $|\cos\theta_{\tau^-}|$.
 This  provides
the product    $[F_i + \lambda G_i][F_j + \lambda G_j]$ as a function of
 $|\cos\theta_{\tau^-}|$ in the Monte Carlo,
 which  uses the fact that the detector is symmetric in
\cst. As a consequence, the forward and backward
hemispheres  use the same Monte Carlo sample. Therefore, the
 correlations in the Monte Carlo samples result in a reduced Monte
 Carlo statistical error on  \aplfb .

The Monte Carlo statistics are taken into account in the likelihood fit in the manner 
described in Reference~\cite{Barlow}.
In order to identify the contribution to the total error
arising from the data statistical error only,
a second fit is performed which does not take into account the
Monte Carlo statistical errors. The  Monte Carlo statistical error 
is taken to be the quadratic difference between
the error from the two fits and
 quoted as part of the systematic error of the polarization results. 

 The effects on the measured polarization arising
from misidentified $\tau$ decays are modelled by the Monte Carlo
simulation. The helicity dependence of the misidentified decays
is automatically taken into account in the product
$[F_i + \lambda G_i][F_j + \lambda G_j]$.
Contributions from the small non-$\tau$ background are estimated
using Monte Carlo simulations of distributions in the relevant kinematic
variables. As there is no helicity dependence in this background,
these distributions are added to the linear combination 
of the right-handed and left-handed $\tau$ decay Monte Carlo
distributions to form the complete reference distributions used in the
fit.

The fit also depends on $\afb$ for which the measured value
in the \Z\ra\tautau ~channel~\cite{bib-z0par} at the appropriate
\ecm ~is used.  Separate distributions for the different 
values of \ecm ~are used in order to account for the \afb ~dependence
but a single fit for \pta ~and \aplfb ~is performed. Although there
are potential dependences of the observables in the analysis on the exact
value of \ecm ~at which the data were collected,
 the use of beam-energy normalized observables
 renders the analysis relatively insensitive to such effects. 
However, in order to further reduce any such dependences,
 the data collected with \ecm ~below 90.7~GeV and above 91.7~GeV
 are analysed using  Monte Carlo samples generated at fixed centre-of-mass
 energies where most of the off-peak data were collected.
 The majority of the off-peak data were collected with values of \ecm ~within
 0.2~GeV of the values used in the Monte Carlo generation.

\section{Polarization Fit Results}
The  results of the global fit are:
\bea
\label{eq-Polresult}
\pta   & = & (\PTALL \pm \PTALLST  \pm \PTALLSY  )\% \nonumber \\
\aplfb  & = & (\APFBALL \pm \APFBALLST  \pm \APFBALLSY  )\% ,
\eea
where the first error is statistical and the second systematic.
The  correlation between the two parameters, including both statistical
and systematic correlations, is  less than  0.03. 
Although this result uses data  collected over a number of different
centre-of-mass energies, to a very good approximation it
can be treated as though it were all collected at a single effective
centre-of-mass energy of 91.30~GeV. Whereas the 
systematic error evaluation is discussed in detail in 
Sections~\ref{sec-detector-systematics} and \ref{sec-physics-systematics},
this section will outline a number of studies which validate the internal
 consistency of the results.

The global fit technique has been checked with independent 
fits to each channel, the results of which are presented
 in Table~\ref{table-channelresults}. The weighted average of
\pta ~for these fit results differs slightly from that obtained from the
global fit. This difference
 is consistent with expected statistical fluctuations when comparing
results obtained using the global
 fit which takes into account
correlations and a weighted average which does not.
 The values obtained
for the five different channels are also consistent with each other and the 
global fit values.
 A $\chi^2$ of 4.9 for four degrees of freedom is found
 when comparing the five values of  \pta ~to the  value from the 
global fit and  2.1 for four degrees of freedom when comparing the
 values of \aplfb.

\begin {table}[htbp]
\begin{center}
\begin{tabular}{|l|c|c|c|c|c|} \hline
                 & \tel    & \tmu    & \tpi   & \tro      & \taone     \\ \hline
 Sample size     & \Nel    & \Nmu    & \Npi    & \Nrho     & \Naone     \\
 Efficiency      & \eleff  & \mueff  & \pieff  & \rhoeff   & \aoneeff   \\
 Background      & \elbkg  & \mubkg  & \pibkg  & \rhobkg   & \aonebkg   \\ \hline
 \pta  (\%)      &\ptauel  &\ptaumu  & \ptaupi &\ptaurho   & \ptauaone  \\
 \aplfb (\%)     &\apolfbel&\apolfbmu&\apolfbpi&\apolfbrho & \apolfbaone\\
\hline 
\end {tabular}
\caption[]{The  number of decays in the sample,
  selection efficiency after tau pair selection
  within the fiducial acceptance and  background
  for each decay mode analysed. Results of independent fits for the
 individual decay modes are also presented where the error quoted 
 represents that arising from the data statistics only. The measurements from 
 the individual channels are correlated and therefore should not be combined in a 
 simple average.
}
\label{table-channelresults}
\end{center}
\end{table}

\begin{figure} 
  \begin{center}
  \mbox{\epsfig{file=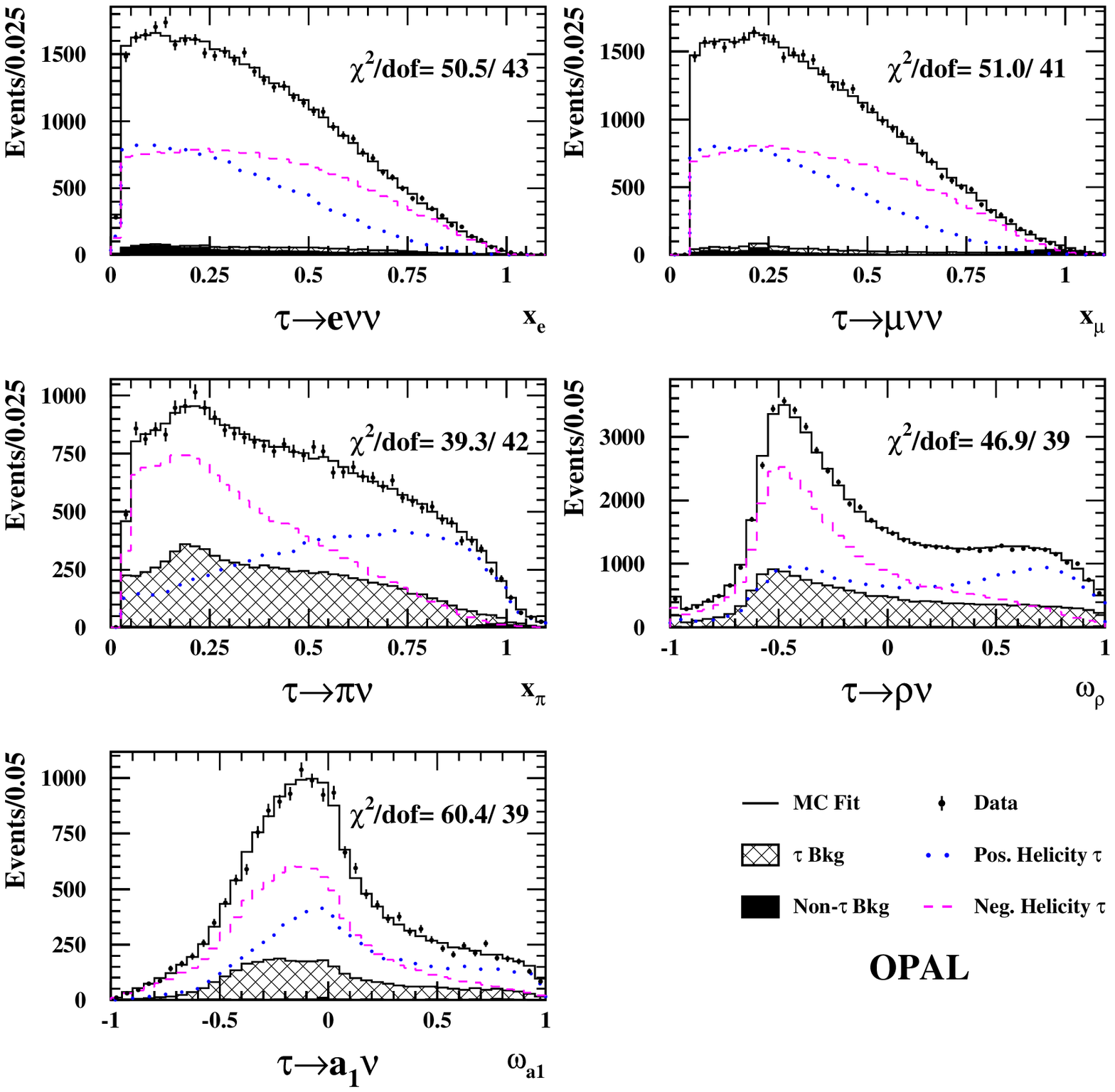,width=18cm,height=20cm}}
 \end{center}
\caption[]{\small Distributions in the
kinematic variables  used in the fits as discussed in the text 
for the  \tel ,  \tmu ,  \tpi,  \tro, and  \taone
 ~channels where the data, shown by points with error bars,
 are integrated over the whole \cst ~range.
Overlaying these distributions are Monte Carlo distributions for
the positive (dotted line) and negative (dashed line) helicity
 $\tau$ leptons and for their sum including background,
assuming a value of \pta = $\PTALL$\% ~as reported in the text.
The hatched histogram represent the  Monte Carlo
expectations of contributions from cross-contamination
 from other $\tau$ decays and the
dark shaded histogram the background from non-$\tau$ sources.
The level of  agreement 
between the data and Monte Carlo distributions is quantified
by quoting the  $\chi^2$  and the number of degrees of freedom.
}
{ \label{fig-xdistrib} }
\end{figure}

An indication of  the validity of the Monte Carlo simulation of the
kinematic variables  used in the  fit, as well as the assumed efficiencies and
purities, is provided by the one-dimensional distributions of the relevant
kinematic variables for the five channels. The distributions 
  combined over all \cst ~bins are shown in
Figure~\ref{fig-xdistrib} for both the data and Monte Carlo.
The $\chi^2$ comparison between data and
 Monte Carlo simulation for these distributions, shown on each plot,
 suggests that the efficiencies and background are adequately simulated
 in these distributions for all modes. 
Also shown are the Monte Carlo distributions of
the variables for  positive and negative helicity $\tau$ lepton decays
 and their sum
including non-$\tau$ background, assuming the value of \pta ~quoted
in Equation~\ref{eq-Polresult}. These plots illustrate the
 sensitivities of the measured
kinematic distributions to the polarization.

\begin{figure} 
  \begin{center}
 \mbox{\epsfig{file=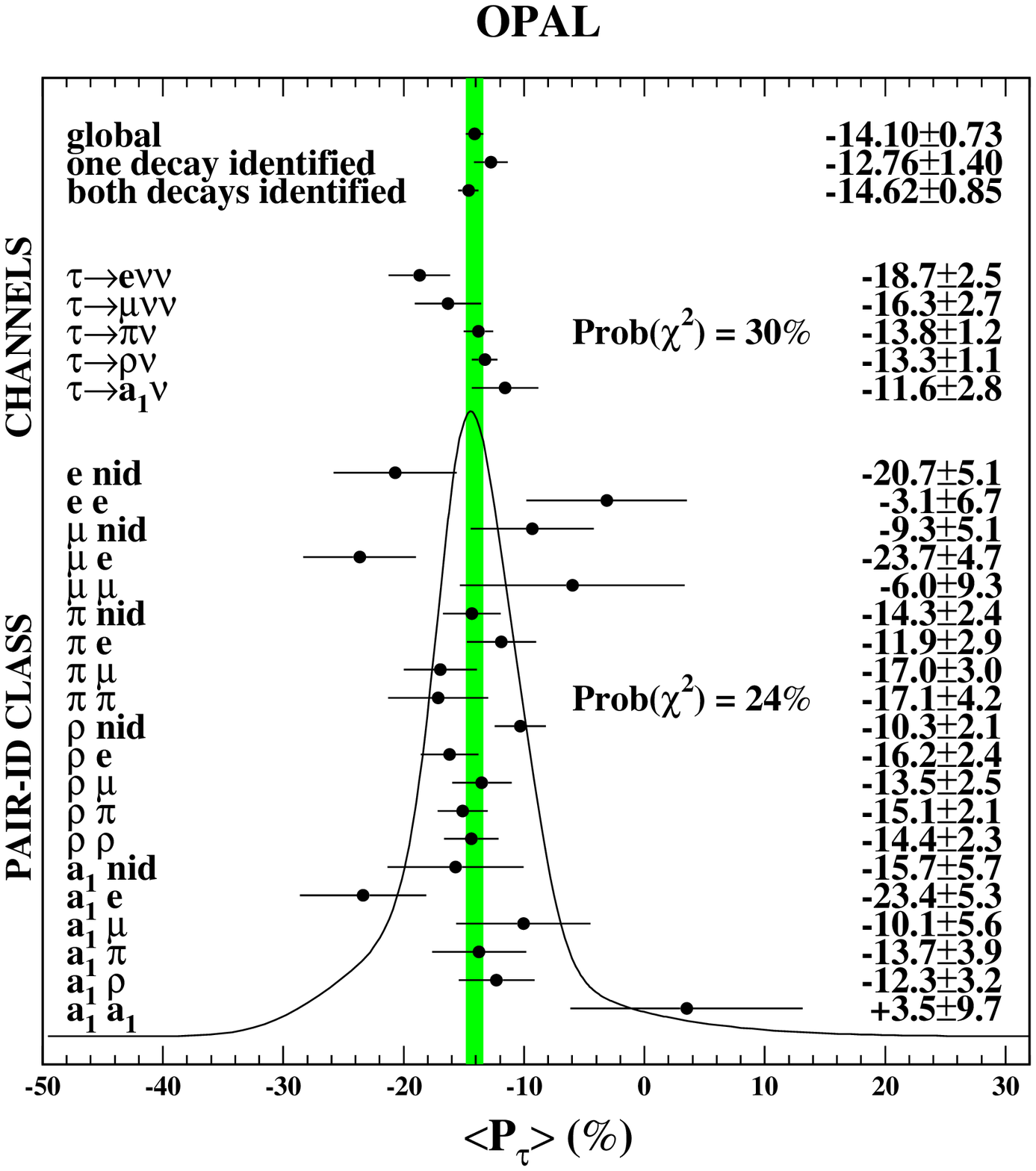,width=18.cm,height=18.cm}}
 \end{center}
\caption[]{\small
Internal consistency of the \pta ~results investigated as a function of
the number of $\tau$ decays classified in the event
 and by pair-identification class.
The ideogram formed from the sum of the individual 
 Gaussians is superimposed on the pair-identification results.
The $\chi^2$ probabilities of the spreads about the global fit value are
shown for each subsample and show good internal consistency in all cases.
 The label `nid' refers to the case where the $\tn$ decay is not
 identified.
}
{ \label{fig-topoconsistpta} }
\end{figure}

\begin{figure} 
  \begin{center}
 \mbox{\epsfig{file=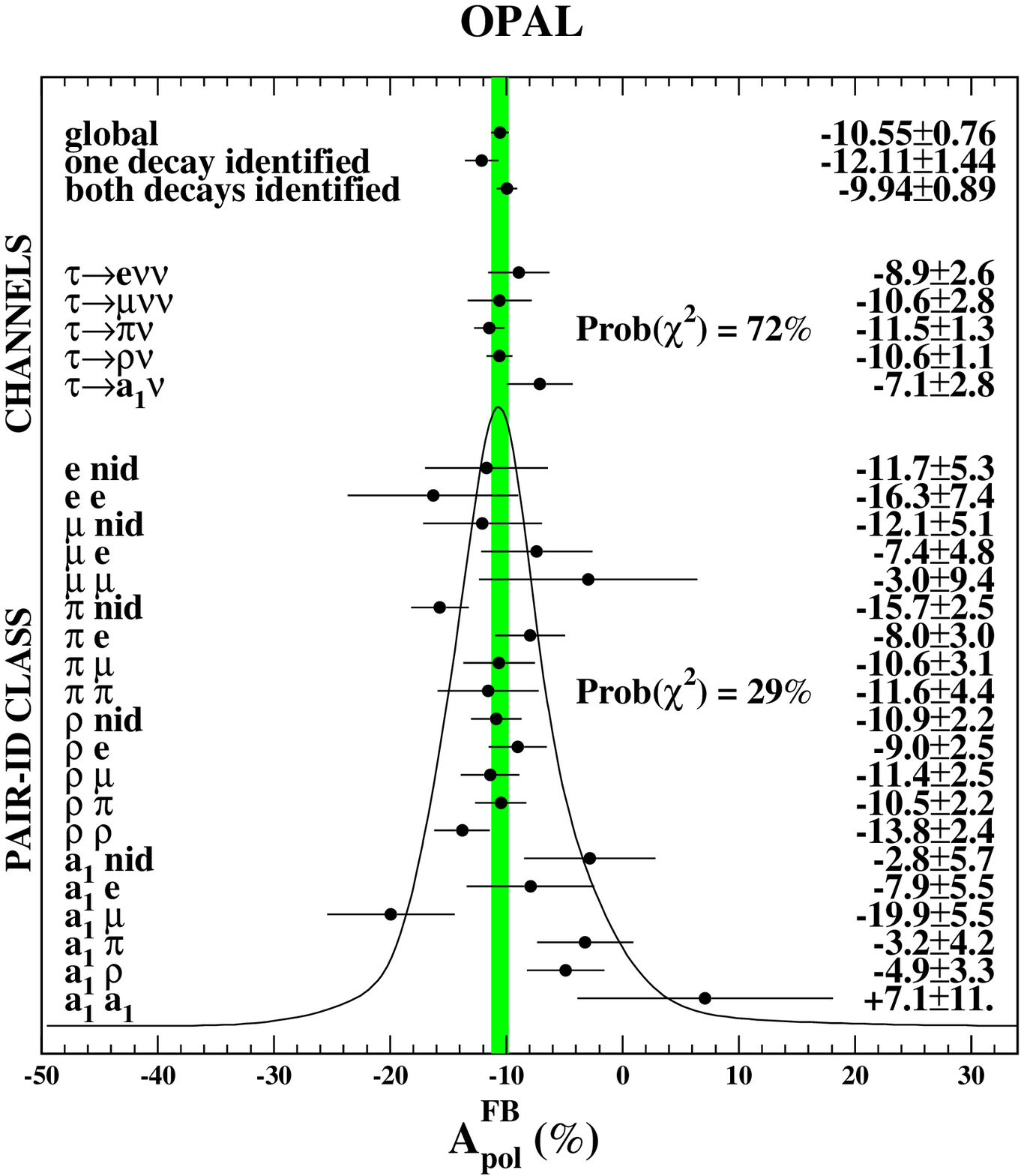,width=18.cm,height=18.cm}}
 \end{center}
\caption[]{\small
Internal consistency of the \aplfb ~results investigated as a function of
the number of $\tau$ decays classified in the event
 and by pair-identification class.
The ideogram formed from the sum of the individual 
 Gaussians is superimposed on the pair-identification class results.
The $\chi^2$ probabilities of the spreads about the global fit value are
shown for each subsample and show good internal consistency in all cases.
 The label `nid' refers to the case where the $\tn$ decay is not
 identified.
}
{ \label{fig-topoconsistaplfb} }
\end{figure}

Another tool for investigating the internal consistency of the analysis is
afforded by comparing the fitted values of \pta ~and \aplfb ~for each event
pair-identification class as defined in Section~\ref{subsec-topo}.
 These comparisons are shown in
 Figures~\ref{fig-topoconsistpta} and~\ref{fig-topoconsistaplfb}
where the results of the fits both in graphical and numeric form for
each pair-identification class in the global analysis are presented.
The $\chi^2$ probability describing the statistical significance of
 the different 
 values from the global fit value\footnote{These are calculated
 using data statitistical errors only.} for \pta ~and \aplfb ~indicate 
 internal consistency. The ideogram formed from the sum of the individual 
 Gaussians is overlayed on the data points. This illustrates that for both 
 \pta ~and \aplfb ~the spread in fit values for the
 different pair-identification classes
 is symmetric about the global fit value and the peak is consistent with that value.

 High statistics internal consistency is also examined
 for events which have both $\tau$
 decays classified compared to those where only one $\tau$ 
 decay is classified.  These comparisons, also shown in 
 Figures~\ref{fig-topoconsistpta} and~\ref{fig-topoconsistaplfb},
 indicate strong overall internal consistency of the results.

\begin{figure}[ht] 
  \begin{center}
 \mbox{\epsfig{file=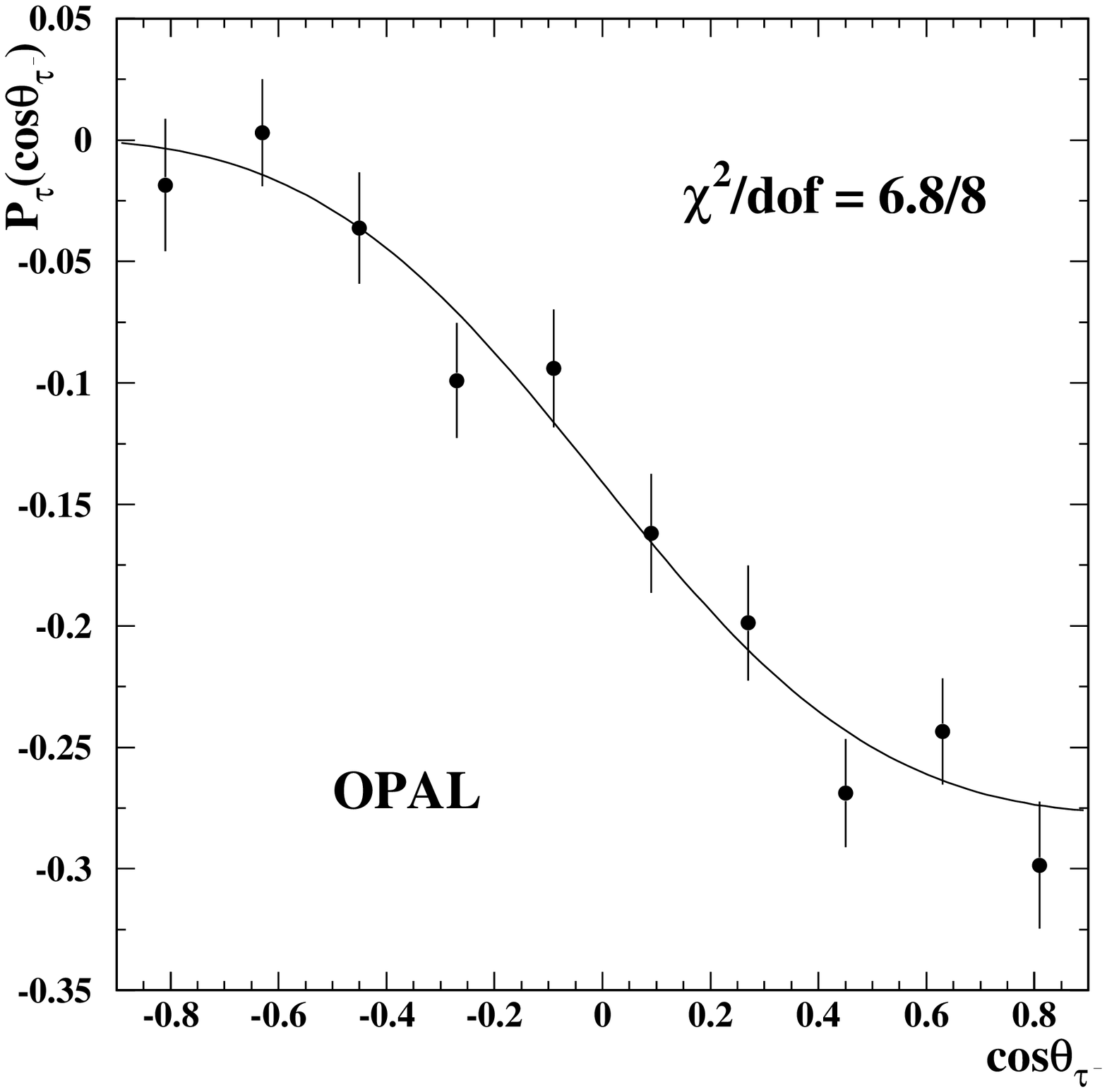,width=13.cm,height=13.cm}}
 \end{center}
\caption[]{\small
Tau polarization, \pta , as a function of $\cst$.
The data points represent the \ptau ~values obtained
from a global fit to all channels in each $\cst$ bin.
The error bars represent data statistical errors only.
The  curve represents the expectation from the global fit result:
 \pta = $\PTALL$\% ~and \aplfb = $\APFBALL$\% as reported in the text.}
{ \label{fig-ALLcos} }
\end{figure}

 As a further check on the validity of the fit,
the results of fits for \ptau 
~performed independently in ten bins of \cst ~are  shown in 
Figure~\ref{fig-ALLcos}. For the fit in a particular
 $\cos\theta_{\tau^-}$ bin,
 an expression analogous to that shown in Equation~\ref{dsig3} is
 used in which 
$(\pta (1+\cstsq)+  \eitovth \aplfb\cst)$ in the second line of
 Equation~\ref{dsig3} is replaced by
 $(\ptau (1 + \cstsq + \eitovth \afb\cst))$.
 This substitution uses Equation~\ref{eq-ptcos}.
Overlaying these points is a curve which 
represents the expectation value of \ptau ~as a function of \cst
~from Equation~\ref{eq-ptcos}
where the values of the \pta ~and \aplfb ~from the global fit of
 Equation~\ref{eq-Polresult} are used.
The  results of the ten independent fits
 are in good agreement with the expectations from the global fit:
the $\chi^2$ is 6.8 for eight degrees of freedom when comparing the
ten values of  P$_{\tau}$ to the expected  value from the global fit where
only the data statistics are included in calculating 
the $\chi^2$.

Because a non-negligible amount of data is collected off-peak, measurements
of \pta ~and \aplfb ~are also performed separately for data collected
at centre-of-mass energies below, on and above the \Z ~resonance peak.
The three statistically independent measurements are quoted
in Table~\ref{table-Ptaucm}. In order to compare the consistency between these
 measurements, they are all converted into measurements 
 of \Atau ~or \Aele ~using the technique described in
 Section~\ref{sec-AsymmResults}.
 These asymmetry values, also quoted in Table~\ref{table-Ptaucm},
 are consistent with each other.
 Because the conversions from \pta ~to \Atau  ~and  from \aplfb to \Aele
~assume the \SM ~centre-of-mass dependence of \pta ~and \aplfb , 
the agreement between the \Atau ~and \Aele ~values in
Table~\ref{table-Ptaucm} indicates that the data are consistent with
 the \SM ~expectations for this centre-of-mass dependence.

\begin {table}[htbp]
\begin{center}
\begin{small}
\begin{tabular}{|l|c c|c c|} \hline
     \ecm        & \pta   & \aplfb     &  \Atau   & \Aele    \\ 
     (GeV)       &  (\%)  &  (\%)      &          &         \\ \hline
     \pminus     & \ptam  & \aplfbm    &  \atm    & \aem    \\
     \peak       & \ptapk & \aplfbpk   &  \atpk   & \aepk    \\        
     \pplus      & \ptap  & \aplfbp    &  \atp    & \aep    \\        
\hline 
\end {tabular}
\end{small}
\caption[]{
Global fit values of \pta ~and \aplfb ~for data collected below, on and above
the \Z ~resonance peak. The luminosity weighted values of \ecm ~are
quoted in the first column where the error reflects the spread in \ecm ~values
of the data combined in each fit.
 The neutral current asymmetry parameters with their statistical errors,
based on the data collected at the different centre-of-mass energies,
are also quoted.}
\label{table-Ptaucm}
\end{center}
\end{table}

\vspace{12pt}
\noindent
\section{ Detector Related Systematic Errors}
\label{sec-detector-systematics}
Because the Monte Carlo simulation provides the positive and negative
helicity reference distributions in the fit and is used for the
likelihood selections, it is necessary that
the detector response be accurately modelled and 
a systematic error attributed to each relevant
discrepency between the detector simulation and data.
The approach taken in this analysis is to determine the accuracy
of the detector modelling of specific measured quantities, to 
vary that quantity in the simulation, and then to propogate the
influence of that variation through a complete analysis of all
decay modes.

High purity control samples of muons with momenta of approximately
45~GeV from \eemm ~events
are used to determine corrections to the
simulation of the momentum scale and resolution
of the central tracking detector. The systematic uncertainties of these
corrections yield a transverse momentum scale uncertainty of
0.14\% in the barrel region of the detector with higher values of 0.8\% and
0.4\% in the overlap and endcap regions, respectively.
 Studies of measurements of the mass
of the \ksh ~from \ksh\ra\pip\pim ~in $\ee\ra\qq$ hadronic events
provide a calibration point of the momentum scale at the lower energies.
  The tracking resolution corrections are cross-checked
 at lower energies using the transverse momentum distributions
in \eeeemm ~two-photon processes. Modelling uncertainties of the $\cos\theta$
measurements of the tracks in the tracking chambers 
are  evaluated independently of transverse momentum  using back-to-back
tracks from \eemm ~events and cross-checked with other detectors such as the 
presampler. Pure samples of electrons with energies  of approximately
45~GeV from \eeee ~events are used to determine corrections to the
simulation of the energy scale and resolution of the ECAL.
 These corrections were evolved to
 lower energies using the ratio of the deposited energy
to measured momentum for electrons in \eeeeee ~two-photon processes
and in high purity \tel ~samples. Uncertainties of 0.3\% on the ECAL energy
scale in the barrel, 0.6\% in the overlap and 0.4\% in the endcap are
 estimated from these studies.
The one standard deviation errors on the energy scale 
and momentum scale are used in assessing the
systematic errors on \pta ~and \aplfb  ~from an 
analysis using rescaled energy and momenta which takes into account
the correlations between channels.
 In a similar manner, systematic errors associated with
uncertainties in the parameters used to describe the resolutions of the
ECAL and tracking detector resolution are also assigned.

High purity muon and \tro ~samples are  used to correct the modelling of
the response of the HCAL and muon chambers to muons and hadrons.
Variation of the magnitude of these corrections is used to
assess the systematic error on \pta ~and \aplfb ~associated with
this  modelling. 
Correct modelling of the dE/dx measurement is achieved by
studying the response of the tracking detector to high purity \tel ~and \tmu
~samples selected without using dE/dx information. The 
corrections applied to the dE/dx simulation are changed  in order to 
assess the sensitivity of \pta ~and \aplfb ~to this modelling.
The effects of uncertainties in the amount of material in the central
detector, which potentially affects the photon 
conversion background in the \aone ~channel, were studied and found 
to have a negligible influence on the polarization measurement.

The uncertainty in the modelling of the lateral spread of the electromagnetic
 and hadronic showers in the ECAL contributes 
significantly to the overall systematic error.
This is particularly relevant for the separation of the \tpi ~and
\tro ~samples.
 The influence of these
uncertainties on the polarization measurement is estimated by varying
the thresholds in the cluster definitions for the simulation
and from the cluster position and position resolution uncertainties.
The ECAL cluster position resolution is also sensitive to the 
lateral shower spread in the ECAL. 
A control sample of electrons is used 
to improve the modelling of the ECAL cluster position resolution and to
assign uncertainties to this modelling by varying the magnitude
 of the corrections applied to the simulation.
 Further checks of this class of systematic error were performed  by studying the
stability of the results from the
likelihood selection when excluding individual observables
 related to showering.

There is also a potential systematic error on \aplfb ~related to
charge mis-assignment, which in OPAL is negligible.

 The contributions to the uncertainty on \pta ~and \aplfb ~from these
 various sources are shown
in Table~\ref{table-systematics}. These are obtained by varying the 
associated detector-level quantities, such as corrections to the simulated
 momentum  or HCAL layers hit, and redoing the analysis from the 
tau pair selection,
through the calculation of physics observables such as m$_{\rho}$ and decay
classification, to the global fit for \pta ~and \aplfb.  
The table includes the systematic errors for each of the five channels in addition
to the errors for the global analysis. The systematic correlations between channels
are automatically accounted for in this analysis and fully incorporated into
the systematic error quoted for the global fit results.

\begin {table}[htbp]
\begin{center}
\begin{tabular}{|l|*{6}{cc|}} \hline
 & \multicolumn{12}{|c|}{$\Delta$\pta\ and $\Delta$\aplfb\ } \\ 
\cline{2-13}
 &
   \multicolumn{2}{|c|}{e}        &
   \multicolumn{2}{|c|}{$\mu$}    &
   \multicolumn{2}{|c|}{$\pi$} &
   \multicolumn{2}{|c|}{$\rho$}   &
   \multicolumn{2}{|c|}{\aone}    &
   \multicolumn{2}{|c|}{Global fit} \\ \hline
 Momentum scale/resolution   &
     0.4 &  0.2     &             
     2.1  &  0.1      &             
     0.8  &  0.1      &             
     0.3  &  0.1      &             
     0.4  &  0.2      &             
     0.24  & 0.13      \\            
 ECAL scale/resolution &
     3.2  &  0.1      &             
     0.2  &  0.1      &             
     0.2  &  --       &             
     1.1  &  0.2      &             
     0.3  &  0.1      &             
     0.17 &  0.11      \\            
 HCAL/MUON modelling           &
     0.1  &  --       &             
     1.1  &  0.1      &             
     0.5  &  0.1      &             
      --  &  --       &             
      --  &  --       &             
     0.13  &  0.05       \\            
 dE/dx errors         &
     0.6  &  0.1      &             
     0.3  &  0.1      &             
     0.3  &  0.1      &             
     0.1  &  0.1       &             
     0.3  &  0.1       &             
     0.12 &  0.08       \\            
 Shower modelling in ECAL &
     0.6  &  0.1      &             
     0.2  &  0.1      &             
     0.4  &  0.1      &             
     0.5  &  0.2      &             
     0.4  &  0.1      &             
     0.25  &  0.10      \\            
 Branching ratios     &
     0.1  &  --       &             
     0.1  &  --       &             
     0.2  &  --       &             
     0.2  &  --       &             
     0.2  &  0.1      &             
     0.11  & 0.02       \\            
 \taone ~modelling               &
      --  &  --       &             
      --  &  --       &             
      --  &  --       &             
     0.4  &  --       &             
     0.5  &  0.1      &             
     0.22  & 0.02       \\            
 $\tn\ra 3\pi \geq 1\piz\nut$  modelling   &
     --   &  --       &             
     --   &  --       &             
      --  &  --       &             
      --  &  --       &             
     1.2  &  0.1      &             
     0.11  & 0.04    \\            
 \afb                 &
    --    &   0.2    &             
    --    &  --      &             
    --    &  --      &             
    --    &  --      &             
    --    &  --      &             
    0.03  &  0.02     \\            
 Decay radiation            &
      --  &  --       &             
      --  &  --       &             
      --  &  --       &             
      --  &  --       &             
     0.1  &  --       &             
     0.01 & 0.01       \\            
 Monte Carlo statistics  &
     0.7  &  0.2      &             
     0.8  &  0.3      &             
     0.3  &  0.1      &             
     0.3  &  0.1      &             
     0.8  &  0.2      &             
     0.22  &  0.10      \\    \hline  
 total                &
     3.4  &  0.4      &             
     2.6  &  0.4      &             
     1.2  &  0.2      &             
     1.3  &  0.3      &             
     1.7  &  0.3      &             
     0.55 &  0.25     \\           
\hline
\end {tabular}
\caption[]{Tabulation of systematic errors contributing to
 \pta\ and \aplfb\, when these asymmetries
 are expressed as a percentage, for each of the five
 decay modes analysed and the global fit.
 In each column the error on \pta\ is given first followed by that
 on \aplfb. Systematic error correlations between the five channels are fully
incorporated into the systematic  error on the global result.
 In the second to sixth columns a dash indicates that the listed effect
 contributes less than  0.05\%.
}
\label{table-systematics}
\end{center}
\end{table}

\vspace{12pt}
\noindent
\section{ Physics Related Systematic Errors}
\label{sec-physics-systematics}
 Another class of systematic uncertainties
 relates to our knowledge of $\tau$ production and decay.
 In this category are the errors on measured branching ratios of the
 different $\tau$ decay modes. The branching ratios used are
 obtained from the  averages of the measurements
 in Reference~\cite{bib-PDG2000}. The error
 on \pta  ~and \aplfb  ~associated with the uncertainty of 
 each branching ratio is
 estimated by varying the value used in the global analysis by
 $\pm 1$~standard deviation about its  average value.
 
 The uncertainty in the  modelling of the \taone ~decay introduces
 systematic  errors both in the
 \taone ~channel and in the \tro ~channel where the \taone ~decays represent
 a significant fraction of the selected decays.
Two contributions to the \taone ~modelling
uncertainty are considered: one being the uncertainty in the mass
and width of the \aone ~as obtained from
Reference~\cite{bib-OPALA1} and the other obtained by comparing two
independent theoretical treatments of the 
\taone ~decay~\cite{bib-Kuhn,bib-IMR,bib-pof}.

In addition to the \taone ~modelling uncertainty, the modelling 
of the  $\tn \ra 3\pi \geq 1 \piz \nut$ decays potentially introduces
 a further uncertainty in the analysis of the \taone ~channel.
 This is studied using the Monte Carlo simulation 
by varying the $\tau^- \ra \omega \pi^-$
contribution to the  $\tn \ra 3\pi  \piz \nut$ mode
 and by flipping the
helicity of the $\tau$ in the $\tn \ra 3\pi >  1 \piz \nut$ decays. 
The latter effect is found to contribute negligibly to the 
systematic error.

 Two  smaller sources of error also fall into this general
 category of systematic error: the error associated with the measured value of 
 \afb ~for \eett ~events which is obtained from 
 Reference~\cite{bib-z0par} and the
 uncertainty of the simulation of radiation. Both initial and final state
radiation are sufficiently well modelled in the Monte Carlo and contribute
negligibly to the systematic uncertainties. The simulation of radiation
 in the decay of the $\tau$  is treated in the same manner as is described in
 Reference~\cite{bib-OPALPL2}. The uncertainty in the modelling of radiation in all
 modes is found to contribute negligibly to the overall systematic error.

 The contributions arising from  the various systematic
 errors are summarized in
 Table~\ref{table-systematics}
 for each of the independent analyses
 and for the global analysis which takes into account the
 correlations between channels.

 Because the fits also depend on our knowledge of the non-$\tau$
 sources of background,
 there are potential systematic errors arising from the uncertainty in the
 production cross-sections. Varying the cross-sections within their errors in
 the reference distributions used in the fits makes negligible changes
 to the polarization results. To cross-check the contributions of non-$\tau$
 background,
 distributions of acoplanarity, acolinearity and total event transverse momentum, which
 would have regions enhanced in the two-photon, \eemm, and \eeee ~events,
 were studied and show no indication of uncontrolled sources of non-tau background.
 The samples of two photon and \eeee ~events are further enhanced by examining
 off-peak data where, again, there is no evidence of problems with these sources.
 Another cross-check is provided by studing the numbers of events in
 the different pair-identification classes,
 where, once more, there is 
 no indication that additional systematic errors
 associated with non-tau backgrounds need to be quoted beyond those included in
 accounting for the detector response related systematic errors.

\vspace{12pt}
\noindent
\section{Neutral Current Asymmetry Parameter Results}
\label{sec-AsymmResults}

The polarization measurements quoted in Equation~\ref{eq-Polresult}
are consistent with our previous
measurements~\cite{bib-OPALPL2} but with a total error that has been reduced
by nearly a factor of two. The results are also consistent with
 those published by other LEP collaborations
~\cite{bib-ALE1PL,bib-DELPHI,bib-L3PL}.

As discussed in the introduction the \SM ~gives
 predictions for $\pta$ and $\aplfb$
in terms of \ecm , the mass and width of the $\Z$, \gvl ~and \gal.
The connection is made through the neutral current asymmetry parameters
as defined in Equation~\ref{lamld}.
The measurements of \pta  ~and \aplfb ~are dominated by
 the \Z propagator but also include small
contributions from  the photon propagator, 
photon-Z$^0$ interference and photonic radiative corrections.
 ZFITTER~\cite{bib-ZFITTER} provides the \ecm ~dependent
non-\Z propagator contributions to \pta  ~and \aplfb ~as well as
 higher-order corrections to Equation~\ref{eq-ptcos} within the context
of the \SM. This allows the measured parameters to be expressed in terms
of \Atau ~and \Aele:
\bea
\Atau &  = & \AATAU \pm \AATAUST 
\pm \AATAUSY , \nonumber \\
\Aele & = & \AAEL \pm \AAELST 
\pm \AAELSY .  \nonumber
\eea

Within the context of the \SM ~these results can be interpreted as
measurements of:
\bea
\vovat &  = & \VOVAT \pm \VOVATSI, \nonumber  \\
\vovae & = & \VOVAE \pm \VOVAESI, \nonumber
\eea
where the statistical and
systematic errors of \Atau ~and \Aele ~are added in quadrature before
calculating the errors on \vovat ~and \vovae.
The agreement between these two values indicates that
the data are consistent with the hypothesis of lepton universality.
This test of lepton universality can be expressed as a measurement of the
ratio of the tau vector to axial vector couplings to the
electron vector to axial vector couplings:
\bea
\frac{\vovat}{\vovae} &  = & 1.00 \pm 0.10 . \nonumber  
\eea
If universality is assumed, these data can be averaged
 and expressed as a single leptonic asymmetry parameter:
\bea
\Alep &= & \AALEP \pm \AALEPSIG . \nonumber
\eea
Using Equations~\ref{lamld} and \ref{vovad}, this result can be expressed as:
\bea
\efswsq &= & \SINWALL \pm \SINWALLSI .\nonumber
\eea
    This measurement of \efswsq ~is of similar precision to other individual
    measurements at LEP using various techniques and is in agreement with the
    value of \efswsq ~obtained from a \SM
    ~fit to all LEP electroweak
    data, including previous measurements of
    the $\tau$ polarization\cite{bib-LEPEW}.
    It is also consistent with a determination of \efswsq 
    ~from a measurement of A$_{\mathrm {LR}}$
  by the SLD collaboration\cite{bib-SLD}.

\section{Combined Lineshape and Asymmetry Results from OPAL}
\label{sec-combined}
The tau polarization results can be combined with the measurements from
the leptonic partial widths and forward-backward asymmetries published
by OPAL\cite{bib-z0par} to provide measurements of \gvl ~and \gal ~for
electrons, muons and $\tau$ leptons separately. While \Atau ~and \Aele ~provide
measurements of \gvl/\gal ~for the $\tau$ and electron,
the leptonic partial widths of the \Z
~provide measurements of $\gvl^2+\gal^2$ for all three lepton flavours.
 The forward-backward asymmetries determine
 (\gve\gvl)/(\gae\gal) for electrons, muons and
$\tau$ leptons yielding additional
information about \gvt/\gat ~and \gve/\gae, and provide the only means of
measuring \gvm/\gam ~at LEP.

\renewcommand{\arraystretch}{1.4}    
\begin{table}                                   
\begin{center}                           
\begin{tabular}{|l|r|}\hline 
 $\MZ \gev      $ & 91.1858 $\pm$ 0.0030\\  
 $\GZ \gev      $ &  2.4948 $\pm$ 0.0041\\ 
 $\shadpol  \nb  $ &  41.501 $\pm$ 0.055 \\ 
 \hline
 $\Ree           $ &  20.902 $\pm$ 0.084 \\   
 $\Rmu           $ &  20.811 $\pm$ 0.058 \\  
 $\Rtau          $ &  20.832 $\pm$ 0.091 \\ 
 \hline
 $\Afbpolee      $ &   0.0089 $\pm$ 0.0044 \\   
 $\Afbpolmumu    $ &   0.0159 $\pm$ 0.0023 \\ 
 $\Afbpoltautau  $ &   0.0145 $\pm$ 0.0030 \\ 
 \hline
 $\Aele$           &   0.1454 $\pm$ 0.0114 \\ 
 $\Atau$           &   0.1456 $\pm$ 0.0095 \\    
\hline                                      
\end{tabular}                           
\caption[Input parmeters to the fits for the leptonic couplings.]
{The first nine parameters are the result of fitting the {\SLP} to the 
measured cross-sections and asymmetries measured by
 OPAL~\protect\cite{bib-z0par}.
The parameters \Aele ~and \Atau ~are the result of the current analysis of
the $\tau$ polarization.}                           
\label{tab-leppar}
\end{center}                           
\end{table}                                   

\begin{table}[htbp]  \begin{center}   
\begin{tabular}{|l|rrrrrrrrrrr|}     
\hline                              
\makebox{\rule[-1.5ex]{0cm}{4.5ex}} 
  & $\MZ$             & $\GZ$             & $\shadpol$       
 & $\Ree$            & $\Rmu$            & $\Rtau$          
 & $\Afbpolee$       & $\Afbpolmumu$     & $\Afbpoltautau$
 & $\Aele$            & $\Atau$    \\
 \hline
 $\MZ$             &$   1.00$ &$   0.05$ &$   0.03$ &$   0.11$ &$   0.00$ &$   0.00$ &$  -0.05$ &$   0.08$ &$   0.06$ &$   0.00$ &$   0.00$ \\             
 $\GZ$             &$   0.05$ &$   1.00$ &$  -0.35$ &$   0.01$ &$   0.02$ &$   0.01$ &$   0.00$ &$   0.00$ &$   0.00$ &$   0.00$ &$   0.00$ \\             
 $\shadpol$        &$   0.03$ &$  -0.35$ &$   1.00$ &$   0.15$ &$   0.22$ &$   0.14$ &$   0.01$ &$   0.01$ &$   0.01$ &$   0.00$ &$   0.00$ \\             
 $\Ree$            &$   0.11$ &$   0.01$ &$   0.15$ &$   1.00$ &$   0.09$ &$   0.04$ &$  -0.20$ &$   0.03$ &$   0.02$ &$   0.00$ &$   0.00$ \\             
 $\Rmu$            &$   0.00$ &$   0.02$ &$   0.22$ &$   0.09$ &$   1.00$ &$   0.06$ &$   0.00$ &$   0.01$ &$   0.00$ &$   0.00$ &$   0.00$ \\             
 $\Rtau$           &$   0.00$ &$   0.01$ &$   0.14$ &$   0.04$ &$   0.06$ &$   1.00$ &$   0.00$ &$   0.00$ &$   0.01$ &$   0.00$ &$   0.00$ \\             
 $\Afbpolee$       &$  -0.05$ &$   0.00$ &$   0.01$ &$  -0.20$ &$   0.00$ &$   0.00$ &$   1.00$ &$  -0.02$ &$  -0.01$ &$   0.00$ &$   0.00$ \\             
 $\Afbpolmumu$     &$   0.08$ &$   0.00$ &$   0.01$ &$   0.03$ &$   0.01$ &$   0.00$ &$  -0.02$ &$   1.00$ &$   0.02$ &$   0.00$ &$   0.00$ \\             
 $\Afbpoltautau$   &$   0.06$ &$   0.00$ &$   0.01$ &$   0.02$ &$   0.00$ &$   0.01$ &$  -0.01$ &$   0.02$ &$   1.00$ &$   0.00$ &$   0.01$ \\             
 $\Aele$           &$   0.00$ &$   0.00$ &$   0.00$ &$   0.00$ &$   0.00$ &$   0.00$ &$   0.00$ &$   0.00$ &$   0.00$ &$   1.00$ &$   0.03$ \\             
 $\Atau$           &$   0.00$ &$   0.00$ &$   0.00$ &$   0.00$ &$   0.00$ &$   0.00$ &$   0.00$ &$   0.00$ &$   0.01$ &$   0.03$ &$   1.00$ \\             
\hline
\end{tabular} 
\end{center}  
\caption[correlation matrix for input to fit for the leptonic couplings] 
{ Error correlation matrix for the 11 parameters entering the 
fit for the leptonic couplings presented in Table~\protect\ref{tab-leppar}.}
\label{tab-lepparc11}  
\end{table}

The OPAL measurements of the hadronic and leptonic  cross-sections and
leptonic forward-backward asymmetries at the \Z ~pole
are summarized in terms of nine {\SLP} as defined in 
Reference~\cite{bib-z0par}:  the mass ($\MZ$), width 
($\GZ$) and  the hadronic pole cross-section ($\shadpol$)
  of the \Z ~resonance; the ratios of the hadronic to leptonic partial 
widths ($\Ree$,$\Rmu$ and $\Rtau$); and the leptonic pole forward-backward
asymmetries ($\Afbpolmumu$,$\Afbpolmumu$ and $\Afbpoltautau$).
The values of these nine parameters, together with the polarization
asymmetries reported here, are displayed in Table~\ref{tab-leppar} and the 
error correlation matrix in Table~\ref{tab-lepparc11}.

\begin{table}[htbp]  \begin{center} 
\renewcommand{\arraystretch}{1.4}  
\begin{tabular}{|l|c|c|c|}      
\hline                          
  & Without lepton  &  With lepton  & {\SM}\\                                                                              
  & universality    &  universality &    prediction \\  
\hline                                                
 $\Aele$       & $  0.1375 \pm 0.0093$  &&\\                                                                              
 $\Amu$       & $  0.154  \pm 0.024$  &&\\                                                                              
 $\Atau$     & $  0.1449 \pm 0.0091$  &&\\                                                                              
 $\Alep$       &
            & $  0.1424 \pm 0.0054$& $  0.1450^{+ 0.0030}_{-0.0084}$  \\                                           
\hline                                                                 
\end{tabular}                                                  
\caption[Results for the coupling parameters]             
{The leptonic neutral current asymmetry
 parameters obtained from a fit to the 
$\Z$ parameters given in Table~\protect\ref{tab-leppar}.   
In the last column is given the value of the parameter
calculated in the context of the {\SM} assuming 
the parameter variations given in the text.}  
\label{tab-al}                                         
\end{center}    
\end{table}                                                                                                                         
\begin{table}[htbp]  \begin{center} 
\begin{tabular}{|l|rrr|}          
\hline                           
 & $\Aele   $ & $\Amu    $ & $\Atau   $ \\
\hline                                   
 $\Aele    $ &$  1.00  $ &$  -0.43 $ &$ -0.09 $ \\ 
 $\Amu    $ &$ -0.43  $ &$   1.00 $ &$  0.04 $ \\     
 $\Atau  $ &$ -0.09  $ &$   0.04 $ &$  1.00 $ \\   
\hline                                           
\end{tabular}                             
\end{center}                     
\caption[Coupling parameter correlation matrix]
{Error correlation matrix for the measurements of 
the leptonic neutral current asymmetry parameters, which are    
presented in Table~\protect\ref{tab-al}. }  
\label{tab-alcor}                
\end{table}                                                                                                                         
Information from the tau polarization and forward-backward 
asymmety measurements can be combined to provide measurements
of the three leptonic neutral current asymmetry parameters, $\Aele$, $\Amu$
and $\Atau$ . These and their error correlation maxtrix 
 are shown in Tables~\ref{tab-al} and \ref{tab-alcor}. The results for 
the three lepton species are consistent with each other and agree well
with the prediction of the \SM
\footnote{The \SM ~calculations require the full specification
of the fundamental \SM ~parameters. The main parameters are the masses
of the \Z boson ($\MZ$), the top quark ($\Mtop$) and the Higgs boson ($\MH$),
and the strong and electromagnetic coupling constants, $\als$ and $\alpha$.
As in Reference~\cite{bib-z0par} the calculation of {\SM} predictions use 
the following values and ranges:
$\MZ \, =  91.1856 \pm 0.0030 \; \mbox{GeV}$,
$\Mtop = 175 \pm 5 \; \mbox{GeV}$,
$\MH \, =  150^{+850}_{-60} \; \mbox{GeV}$,
$\als = 0.119 \pm 0.002$, and
$\alpha(\MZ^2)^{-1} \, = 128.886\pm0.090$ . 
The choice of these parameter values and ranges is discussed in 
Reference~\protect\cite{bib-z0par}.}
which is  also shown in Table~\ref{tab-al}. Assuming lepton unversality,
\bea
\Alep = 0.1424 \pm 0.054 .\nonumber
\eea
This can also be expressed in terms of the effective weak mixing angle:
\bea
\efswsq = \SINWALLOPAL \pm \SINWALLSIOPAL .\nonumber
\eea

\begin{figure}[htbp] 
  \begin{center}
 \mbox{\epsfig{file=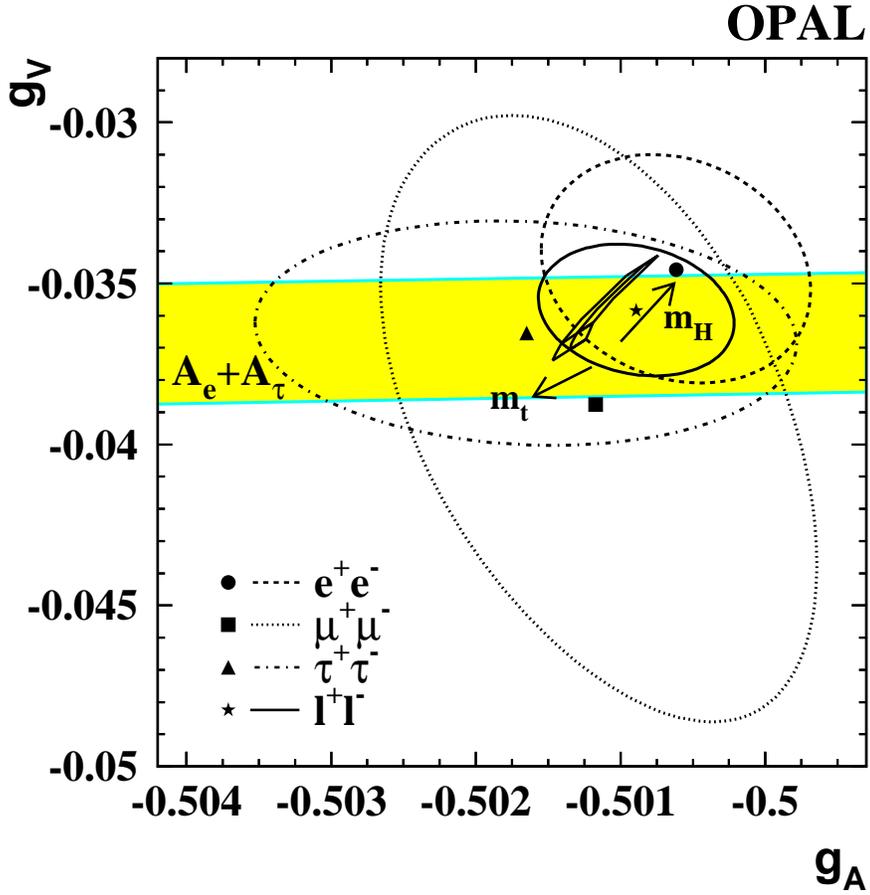,width=13.cm,height=13.cm}}
 \end{center}
\caption[]{\small \gvl ~{\it vs} \gal ~as determined from the
OPAL measurements of the leptonic partial widths of the \Z, 
forward-backward asymmetries and tau polarization measurements.
The ellipses represent the 68\% confidence level contours in the
 \gvl-\gal ~plane
for each lepton species separately (dotted and dashed)
 and for all leptons assuming
 universality (solid). The central values are displayed at the
 centre of the ellipses as a circle, square, triangle and  star
 for electrons, muons, tau leptons and all leptons under universality,
respectively.
 The \SM ~prediction is shown with
variations  from the top quark mass (170 to 180~GeV) and Higgs mass
(90 to 1000~GeV) indicated. 
The OPAL tau polarization measurements of $\Atau$ and $\Aele$ constrain
\gvl ~and \gal ~to lie in the shaded region 
 at the 68\% confidence level.
}
{ \label{fig-gvvsga} }
\end{figure}

The eleven {\SLP} listed in Table~\ref{tab-leppar} can also be used
to determine  the neutral current
 vector and axial-vector couplings for each lepton species.
 The results and the error correlation matrix are given in
 Tables~\ref{tab-gvga} and \ref{tab-gvga6} and are illustrated in 
Figure~\ref{fig-gvvsga}. Some of the couplings have 
 up to approximately 50\% correlations between their errors.
 Evident from these results is the universality
of the coupling constants which can be quantified in 
terms of the ratios of the 
couplings:
\bea
\frac{\gam}{\gae} = 1.0011  \pm 0.0025,&
\frac{\gat}{\gae} = 1.0021  \pm 0.0029,&
\frac{\gat}{\gam} = 1.0009  \pm 0.0027,  \nonumber \\
\frac{\gvm}{\gve} = 1.12 \pm ^{0.24}_{0.21},&
\frac{\gvt}{\gve} = 1.06 \pm ^{0.11}_{0.10},&
\frac{\gvt}{\gvm} = 0.94 \pm ^{0.18}_{0.14} . \nonumber  
\eea
 The axial-vector couplings of the different lepton species
are found to be the same at the 0.3\% level.
The errors on the inter-species ratios of the vector couplings are much
larger because of  the smaller size of the vector couplings themselves, but
within this reduced sensitivity, again no significant differences are
observed.
These errors have been significantly reduced by adding information from the
tau polarization to the lineshape and forward-backward asymmetry
measurements\cite{bib-z0par}.

Combining the values of the coupling constants from the different lepton
species under the assumption of lepton universality yields the values
\bea
\gal = -0.50089  \pm 0.00045, &
\gvl = -0.0358  \pm   0.0014 . \nonumber 
\eea
where the correlation between \gal ~and \gvl ~is $-19\%$, which is
approximately the same as the $-21\%$ correlation between \gal ~and 
\efswsq.
These are in good agreement with the predictions of 
the {\SM}, which are included in Table~\ref{tab-gvga}.

\begin{table}[htbp]  \begin{center}  
\renewcommand{\arraystretch}{1.4}  
\begin{tabular}{|l|c|c|c|}         
\hline                             
  & Without lepton  &  With lepton  & Standard Model\\   
  & universality    &  universality &    prediction \\   
\hline                                                 
 $\gae$       & $  -0.50062 \pm 0.00062$  &&\\       
 $\gam$       & $  -0.50117 \pm 0.00099$  &&\\    
 $\gat$     & $  -0.50165 \pm 0.00124$  &&\\  
 $\gal$       &
           & $  -0.50089 \pm 0.00045$& $  -0.50130^{+ 0.00047}_{-0.00013}$  \\ 
 $\gve$       & $   -0.0346 \pm 0.0023$  &&\\                                                                              
 $\gvm$       & $   -0.0388^{+  0.0060}_{ -0.0064}$  &&\\          
 $\gvt$     & $   -0.0365 \pm 0.0023$  &&\\     
 $\gvl$       &
           & $  -0.0358 \pm 0.0014$& $  -0.0365^{+ 0.0022}_{-0.0008}$  \\                                                 
\hline                    
\end{tabular}            
\caption[Results for the axial and vector couplings] 
{Axial-vector and vector couplings obtained from a fit to the parameter set
given in Table~\protect\ref{tab-leppar}.                                      
In the last column we give the values of the couplings
calculated in the context of the {\SM} assuming        
the parameter variations given in the text. } 
\label{tab-gvga}                                   
\end{center}                                      
\end{table}                                                                               

\begin{table}[htbp]  \begin{center}
\begin{tabular}{|l|rrrrrr|}      
\hline                          
\makebox{\rule[-1.5ex]{0cm}{3.5ex}} 
& $\gae$    & $\gam$       & $\gat$     & $\gve$       & $\gvm$  & $\gvt$ \\
 \hline
 $\gae$       &$ 1.00 $&$ -.17 $&$ -.13 $&$ -.19 $&$  .07 $&$  .01 $ \\
 $\gam$       &$ -.17 $&$ 1.00 $&$  .29 $&$  .19 $&$ -.46 $&$ -.03 $ \\
 $\gat$     &$ -.13 $&$  .29 $&$ 1.00 $&$ -.04 $&$  .03 $&$ -.08 $ \\
 $\gve$       &$ -.19 $&$  .19 $&$ -.04 $&$ 1.00 $&$ -.45 $&$ -.04 $ \\
 $\gvm$       &$  .07 $&$ -.46 $&$  .03 $&$ -.45 $&$ 1.00 $&$  .03 $ \\
 $\gvt$     &$  .01 $&$ -.03 $&$ -.08 $&$ -.04 $&$  .03 $&$ 1.00 $ \\
\hline
\end{tabular}  
\end{center}   
\caption[6x6 leptonic coupling correlation matrix]
{ Error correlation matrix for the measurements of  
the axial vector and vector couplings,           
without assuming lepton                       
universality, which are presented in Table~\protect\ref{tab-gvga}.
}
\label{tab-gvga6}                     
\end{table}                              

\section{Summary}
\label{sec-summary}
Measurements of \pta ~and \aplfb ~have been made using
the complete LEP~I data sample of OPAL. The results are based
 on a simultaneous analysis of  \tel , \tmu , \tpi , \tro ~and \taone
~decays from a sample of \Ntpair ~\eett ~candidates collected over nearly
the entire solid angle of the OPAL detector.
Under the assumption that the $\tau$ lepton decays
according to V$-$A theory,  the average \tn ~polarization
near $\sqrt{s}$~=~\mz ~is measured to be
$\pta= (\PTALL \pm \PTALLST \pm \PTALLSY)\%$
 and the \tn ~polarization forward-backward
asymmetry to be $\aplfb=(\APFBALL \pm \APFBALLST \pm \APFBALLSY)\%$.
Taking into account the small effects of  the photon propagator, 
photon-Z$^0$ interference and photonic radiative corrections, these results
can be expressed in terms of the lepton neutral current asymmetry parameters:
\bea
\Atau &  = & \AATAU \pm \AATAUST 
\pm \AATAUSY , \nonumber \\
\Aele & = & \AAEL \pm \AAELST 
\pm \AAELSY .  \nonumber
\eea
These measurements are
consistent with the hypothesis of lepton universality and
combine to give $\Alep = \AALEP \pm \AALEPSIG$.
 Within the context of the \SM ~this  corresponds
 to $\efswsq=\SINWALL \pm \SINWALLSI$.

 Combining the information from 
 the tau polarization results with the results of the
 other OPAL neutral
 current measurements yields values for the vector and axial-vector
couplings which are the same for all lepton species and gives
\bea
\gal = -0.50089  \pm 0.00045, &
\gvl = -0.0358  \pm   0.0014 . \nonumber 
\eea
Expressing these results in terms of the electroweak mixing angle 
gives
\bea
\efswsq=\SINWALLOPAL \pm \SINWALLSIOPAL .\nonumber
\eea
 This is consistent with
the \SM ~and with the current world average value\cite{bib-LEPEW}.

\bigskip\bigskip\bigskip
\appendix
\par
\noindent
{\large\bf Acknowledgements}\\
\par
\noindent
We particularly wish to thank the SL Division for the efficient operation
of the LEP accelerator at all energies
 and for their continuing close cooperation with
our experimental group.  We thank our colleagues from CEA, DAPNIA/SPP,
CE-Saclay for their efforts over the years on the time-of-flight and trigger
systems which we continue to use.  In addition to the support staff at our own
institutions we are pleased to acknowledge the  \\
Department of Energy, USA, \\
National Science Foundation, USA, \\
Particle Physics and Astronomy Research Council, UK, \\
Natural Sciences and Engineering Research Council, Canada, \\
Israel Science Foundation, administered by the Israel
Academy of Science and Humanities, \\
Minerva Gesellschaft, \\
Benoziyo Center for High Energy Physics,\\
Japanese Ministry of Education, Science and Culture (the
Monbusho) and a grant under the Monbusho International
Science Research Program,\\
Japanese Society for the Promotion of Science (JSPS),\\
German Israeli Bi-national Science Foundation (GIF), \\
Bundesministerium f\"ur Bildung und Forschung, Germany, \\
National Research Council of Canada, \\
Research Corporation, USA,\\
Hungarian Foundation for Scientific Research, OTKA T-029328, 
T023793 and OTKA F-023259.\\


\begin{thebibliography}{99}



\bibitem{bib-PCP} D.~Bardin, M.~Grunewald and G.~Passarino,
``Precision calculation project report,''
hep-ph/9902452.

\bibitem{bib-Jadach} S.~Jadach and Z.~W\c{a}s in {\it Z Physics at LEP1},
                 CERN 89-08, edited by
                 G. Altarelli \etal , Vol. 1 (1989) 235.

\bibitem{bib-z0par}
OPAL Collab.,  M.Z. Akrawy \etal, Phys. Lett. {\bf B240} (1990) 497; \\
 OPAL Collab.,  G. Alexander \etal, Z. Phys. {\bf C52} (1991) 175;  \\
 OPAL Collab.,  P.D. Acton \etal, Z. Phys. {\bf C58} (1993) 219;    \\
  OPAL Collab., R. Akers \etal,  Z. Phys. {\bf C61} (1994) 19; \\
   OPAL Collab., {\it Precise Determination of the Z Resonance
 Parameters at LEP: `Zedometry'},
CERN-EP-2000-148, Submitted to Eur. Phys. J. C.
 to be published.



\bibitem{bib-opal}
 OPAL Collab., K.~Ahmet {\it et al.},
  Nucl.~Inst.~and Meth. {\bf A305} (1991) 275.

\bibitem{bib-trigger}
 OPAL Collab., K.~Arignon {\it et al.},
  Nucl.~Inst.~and Meth. {\bf A313} (1992) 103. 

\bibitem{bib-OPALPL3} OPAL Collab.,
   G.~Alexander \etal , Z. Phys. {\bf C72} (1996) 365.\\
                 The  OPAL combined results from the 1996 paper are:
                 $\Atau = 0.134\pm0.009\pm0.010$, 
                 $\Aele = 0.129\pm0.014\pm0.005$, 
                 $\efswsq=0.2334\pm0.0012$.  

\bibitem{DAVIER} M.Davier \etal, Phys. Lett. {\bf B306} (1993) 411;\\
             A.Roug\'{e}, Proceeding of the Workshop on Tau Lepton Physics,
         M.Davier and B.Jean-Marie Editors, Edition Fronti\`{e}res (1991) 213.

\bibitem{bib-OPALPL2} OPAL Collab.,
   R.~Akers \etal , Z. Phys. {\bf C65} (1995) 1.


\bibitem{bib-OPALPL1} OPAL Collab.,
   G.~Alexander \etal , Phys. Lett. {\bf B266} (1991) 201.

\bibitem{bib-graham} K. Graham,
`Precision Determination of the Electroweak Mixing Angle  and Test of
Neutral Current Universality from the Tau Polarization Measurements at OPAL',
  University of Victoria PhD Thesis (in preparation). 
  Will be available from the National Archives of Canada.
 
\bibitem{bib-koralz}
      S. Jadach, J.H. K\"{u}hn and Z. W\c{a}s, (TAUOLA) Comp. Phys. Comm.
                 {\bf 64} (1990) 275;\\
S.~Jadach, \etal , 
(TAUOLA 2.4)
Comp.\ Phys.\ Comm.\  {\bf 76} (1993) 361;\\
 S. Jadach, B.F.L Ward and Z. W\c{a}s,  (KORALZ 3.8)
Comp. Phys. Comm.                  {\bf 66} (1991) 276; \\
 S. Jadach, B.F.L Ward and Z. W\c{a}s, (KORALZ 4.0)
Comp. Phys. Comm.  {\bf 79} (1994) 503;\\
S.~Jadach, B.~F.~Ward and Z.~Was, (KORALZ 4.02)
Comp.\ Phys.\ Comm.\  {\bf 124} (2000) 233.\\
Events were generated with  KORALZ 4.02 setting $s'/s>0.00156$
and TAUOLA 2.6 with four pion decays modelled according to:\\
R.~Decker, M.~Finkemeier, P.~Heiliger and H.~H.~Jonsson,
Z.\ Phys.\  {\bf C70} (1996) 247.

\bibitem{bib-bhwide} S. Jadach, W. Placzek and B.F.L. Ward, (BHWIDE)
Phys.Lett. {\bf B390} (1997) 298.

\bibitem{bib-Jetset}  T.~Sj\"{o}strand,
 Comp. Phys. Comm. {\bf 39} (1986) 347;\\
 M.~Bengtsson and T.~Sj\"{o}strand,
 Comp. Phys. Comm. {\bf 43} (1987) 367; \\
 M.~Bengtsson and T.~Sj\"{o}strand,
 Nucl.~Phys. {\bf B289} (1987) 810.

\bibitem{bib-OPALQCD}
 OPAL Collab., P. Acton {\it et al.},
  Z. Phys. {\bf C58} (1993) 387.

\bibitem{bib-vermas}
 R.~Battacharya, J.~Smith and G.~Grammer,  Phys. Rev. {\bf D15} (1977) 3267;\\
 J.~Smith, J.A.M.~Vermaseren and G.~Grammer, Phys. Rev. {\bf D15} (1977) 3280.
 
\bibitem{bib-hadronictwophoton}
(PHOJET 1.05c used with JETSET 7.408): \\
R. Engel and J. Ranft, Phys. Rev.  {\bf D54} (1996) 4244;\\
R. Engel, Z. Phys. {\bf C66} (1995) 203;\\
A.~Buijs, W.~G.~Langeveld, M.~H.~Lehto and D.~J.~Miller, (TWOGEN)
Comp.\ Phys.\ Comm.\  {\bf 79} (1994) 523.

\bibitem{bib-FERMISV}
 J. Hilgart, R. Kleiss, F. Le Diberder,
Comp. Phys. Comm.  {\bf 75} (1993) 191.

\bibitem{bib-gopal} J. Allison \etal, Nucl. Inst. and Meth. {\bf A317}
                 (1992) 47.
 
 
\bibitem{bib-geant}
 R. Brun, F. Bruyant, M. Maire, A.C. McPherson, and P. Zanarini,
 {\it GEANT3}, CERN DD/EE/84-1 (1987).

\bibitem{bib-Karlen} D. Karlen, Computers in Physics, 12:4 (1998) 380.

\bibitem{bib-maxent} M. Thomson, Nucl. Inst. and Meth. {\bf A382} (1996) 553.

\bibitem{bib-a1_3prong} ARGUS Collab., H. Albrecht \etal,
  Z. Phys. {\bf C58} (1993) 61.


\bibitem{bib-Kuhn}
 J.H.~K\"{u}hn and A.~Santamaria  Z. Phys. {\bf C48} (1990) 445;\\
J.H.~K\"uhn and E.\ Mirkes,
Z. Phys. {\bf C56} (1992) 661.

\bibitem{bib-IMR}
N.\ Isgur, C.\ Morningstar and C.\ Reader,
Phys. Rev. {\bf D39} (1989) 1357.

\bibitem{bib-Tsai}
Y.~Tsai,
Phys.\ Rev.\ {\bf D 13} (1976) 771.


\bibitem{Barlow} R. Barlow and C. Beeston, Comp. Phys. Comm. {\bf 77} (1993) 219.

\bibitem{bib-ZFITTER} D.~Bardin \etal, 
``ZFITTER: An Analytical program for fermion pair production
 in e+ e- annihilation'',  CERN-TH. 6443/92 (1992), hep-ph/9412201;
D.~Bardin, P.~Christova, M.~Jack, L.~Kalinovskaya, A.~Olchevski, S.~Riemann and T.~Riemann,
Comput.\ Phys.\ Commun.\ {\bf 133} (2001) 229
[hep-ph/9908433].

\bibitem{bib-PDG2000}
The Particle Data Group, D.E. Groom \etal,
Eur. Phys. J. {\bf C15} (2000) 1.\\
(URL: http://pdg.lbl.gov/).

\bibitem{bib-OPALA1} 
OPAL Collab., R. Akers \etal, Z. Phys. {\bf C67} (1995) 45.


\bibitem{bib-pof}
P.~R.~Poffenberger, Z. Phys.  {\bf C71} (1996) 579.
 
\bibitem{bib-ALE1PL} ALEPH Collab.,
                 D. Decamp \etal, Phys. Lett. {\bf B265} (1991) 430;\\
                 ALEPH Collab.,
                 D. Buskulic \etal, Z. Phys. {\bf C59} (1993) 369;\\
                 ALEPH Collab.,
                 D. Buskulic \etal, Z. Phys. {\bf C69} (1996) 183.\\
                 Using all published ALEPH $\tau$ polarization measurements,
                 ALEPH quotes  combined results of
                 $\Atau = 0.136\pm0.012\pm0.009$, 
                 $\Aele = 0.129\pm0.016\pm0.005$, 
                 $\efswsq=0.2332\pm0.0014$.

\bibitem{bib-DELPHI} DELPHI Collab.,
                 P. Abreu \etal,  Z. Phys. {\bf C55} (1992) 555;\\
                DELPHI Collab., P. Abreu \etal, Z. Phys. {\bf C67} (1995) 183;\\
                 DELPHI Collab., P. Abreu \etal, Eur. Phys.J. {\bf C14} (2000)
                  585-611.\\
                 Using all published DELPHI $\tau$ polarization measurements,
                 DELPHI quotes  combined results of
                 $\Atau = 0.1359\pm0.0079\pm0.0055$, 
                 $\Aele = 0.1382\pm0.0116\pm0.0005$, 
                 $\efswsq=0.23282\pm0.00092$. 

\bibitem{bib-L3PL} L3 Collab.,  O. Adriani \etal, Phys. Lett. {\bf B294} (1992) 466.\\
                   L3 Collab.,  M. Acciarri \etal,  Phys. Lett. {\bf B341} (1994) 245;\\
                   L3 Collab., M.~Acciarri {\it et al.}, 
                   Phys.\ Lett.\ {\bf B429} (1998) 387.\\
                 Using all published L3 $\tau$ polarization measurements,
                 L3 quotes  combined results of
                 $\Atau = 0.1476\pm0.0088\pm0.0062$, 
                 $\Aele = 0.1678\pm0.0127\pm0.0030$, 
                 $\efswsq=0.2306\pm0.0011$.  

\bibitem{bib-LEPEW} LEP Collaborations and the LEP Electroweak Working Group,
        {\em A Combination of Preliminary LEP Electroweak
        Results and Constraints on the \SM},
      CERN-EP-2000-016, January 2000. \\
     All the asymmetry measurements made at LEP, including
       some preliminary results and 
       including the $\tau$ polarization, yields an average value of
     \efswsq=0.23192$\pm$0.00023.
     The \SM ~fit to all LEP electroweak data, including M$_W$ and
     the $\tau$ polarization, yields a value of
      \efswsq=0.23150$\pm$0.00016.

     The LEP average value from the forward-backward asymmetry from all leptons is
     \efswsq=0.23107$\pm$0.00053; from the forward-backward hadronic charge asymmetry is
     \efswsq=0.2321$\pm$0.0010; from the forward-backward asymmetry from  c-quarks is
     \efswsq=0.23255$\pm$0.00086; and from the forward-backward asymmetry from  b-quarks is
     \efswsq=0.23228$\pm$0.00036. 

\bibitem{bib-SLD}
SLD Collab., K. Abe \etal, Phys. Rev. Lett. {\bf 84} (2000) 5945. \\
 This measurement of A$_{\mathrm{LR}}$ yields a value 
 of \efswsq=0.23097$\pm$0.00027. 

\end{thebibliography}
\end{document}